\documentclass[floats,floatfix,showpacs,preprintnumbers,amssymb,prd,twocolumn,superscriptaddress,nofootinbib,nolongbibliography,reprint]{revtex4-1}

\usepackage{amssymb,amsmath,verbatim,mathtools,needspace,enumitem,etoolbox,graphicx,physics,microtype,afterpage,xspace,tabularx,lmodern,multirow,hhline,boldline, makecell, booktabs, subfigure}
\usepackage{gensymb}
\usepackage{appendix}
\usepackage{tensor}
\usepackage{array}
\usepackage{siunitx}
\usepackage[normalem]{ulem}
\usepackage[dvipsnames, usenames]{xcolor}
\usepackage{xr-hyper}
\definecolor{linkcolor}{rgb}{0.0,0.3,0.5}
\usepackage[unicode, colorlinks=true, linkcolor=linkcolor, citecolor=linkcolor, filecolor=linkcolor, urlcolor=linkcolor, linktocpage, breaklinks]{hyperref}
\usepackage[all]{hypcap}
\usepackage[T1]{fontenc}
\usepackage[utf8]{inputenc}
\usepackage[usenames,dvipsnames]{xcolor}
\hypersetup{colorlinks=true,citecolor=magenta,linkcolor=violet,urlcolor=magenta}
\usepackage{multirow}

\definecolor{romared}{RGB}{142,0,28}

\newcommand{\be}{\begin{equation}}
\newcommand{\ee}{\end{equation}}

\def\be{\begin{equation}}
\def\ee{\end{equation}}
\newcommand{\beq}{\begin{eqnarray}}
\newcommand{\eeq}{\end{eqnarray}}

\usepackage{makecell}
\usepackage{soul}

\newcolumntype{Y}{>{\centering\arraybackslash}X}

\makeatletter
\newcommand*{\addFileDependency}[1]{
  \typeout{(#1)}
  \@addtofilelist{#1}
  \IfFileExists{#1}{}{\typeout{No file #1.}}
}
\makeatother

\begin{document}

\title{Gravitational Atoms from Topological Stars}

\author{Ibrahima Bah}
\email[]{iboubah@jhu.edu}
\affiliation{William H. Miller III Department of Physics and Astronomy, Johns Hopkins University, 3400 North Charles Street, Baltimore, Maryland, 21218, USA}

\author{Emanuele Berti}
\email[]{berti@jhu.edu}
\affiliation{William H. Miller III Department of Physics and Astronomy, Johns Hopkins University, 3400 North Charles Street, Baltimore, Maryland, 21218, USA}

\author{Bogdan Ganchev}
\email[]{bganche1@jh.edu}
\affiliation{William H. Miller III Department of Physics and Astronomy, Johns Hopkins University, 3400 North Charles Street, Baltimore, Maryland, 21218, USA}

\author{David Pere\~niguez}
\email[]{dpereni1@jhu.edu}
\affiliation{William H. Miller III Department of Physics and Astronomy, Johns Hopkins
University, 3400 North Charles Street, Baltimore, Maryland, 21218, USA}

\author{Nicholas Speeney}
\email[]{nspeene1@jhu.edu}
\affiliation{William H. Miller III Department of Physics and Astronomy, Johns Hopkins
University, 3400 North Charles Street, Baltimore, Maryland, 21218, USA}

\date{\today}

\begin{abstract}
We study the bound states of a massive scalar field around a topological star, and show that these are strictly normal modes. This yields a genuine gravitational atom, sharply distinguishing horizonless objects from black holes. We show that the modes are controlled by the field's Compton wavelength compared to the size of the star. When the Compton wavelength is large, the field forms a cloud with a hydrogen-like spectrum, while in the opposite regime it is localized along timelike trajectories. When the two scales are comparable the spectrum becomes richer, and we characterize it in detail allowing the field to carry electric charge and Kaluza--Klein momentum.
\end{abstract}

\maketitle

\section{Introduction}

The existence of event horizons in our universe is a robust prediction of general relativity (GR), as follows from combining singularity theorems~\cite{Penrose:1964wq}, cosmic censorship conjectures~\cite{Penrose:1969pc} and the genericity of trapped surfaces~\cite{Schoen:1983tiu,Christodoulou:2008nj}. Gravitational wave (GW) observatories~\cite{LIGOScientific:2016aoc,LIGOScientific:2020ibl} and long baseline interferometry~\cite{EventHorizonTelescope:2019dse,GRAVITY:2020gka} give support to such predictions on the empirical side. However, horizons pose major conceptual challenges such as the formation of spacetime singularities, the entropy problem and the information paradox~\cite{Bena:2022ldq,Carballo-Rubio:2025fnc}, none of them being resolvable by GR on its own. These reasons are strong enough to motivate the exploration of alternatives to the black hole (BH) paradigm, despite the strong theoretical and observational evidence for their existence~\cite{Cardoso:2019rvt,Berti:2025hly}.

Self-gravitating objects that source fields as strong as those of BHs, but that are not cloaked by an event horizon, are challenging to conceive~\cite{Bambi:2025wjx}. More precisely, such objects should exhibit a photon sphere and possess a high-redshift surface in order to mimic a BH robustly. In addition, resolving the horizon problem with BH mimickers requires that not just some, but \textit{all} BHs in the Universe are replaced by horizonless bodies. In particular, this implies that the mass and size of BH mimicker candidates should be free parameters (or cover a wide enough range), in order to fit all observed compact objects in e.g. GW data. Finally, horizonless alternatives to BHs should exist within a well-defined physical theory, where classical dynamics as well as high-energy completions can be studied in a consistent framework and yield sharp predictions. 

An interesting proposal satisfying the requirements above builds on one of our current paradigms in quantum gravity, and consists in understanding BHs as ensembles of quantum states. As is often the case with quantum systems, there are regions of parameter space where coherent subsets of these states are susceptible to a classical description. Indeed, several families of smooth, horizonless geometries have been constructed in well-defined theories~\cite{Bena:2015bea,Bena:2016agb,Bena:2016ypk,Bena:2017xbt,Ceplak:2018pws}, many of which are believed to correspond to concrete BH microstates~\cite{Giusto:2004id,Giusto:2012yz,Giusto:2015dfa,Galliani:2016cai,Giusto:2020mup,Ceplak:2021wzz,Rawash:2021pik,Ganchev:2021ewa}.  Most of these constructions have been carried out in a supersymmetric setting, but in recent years a growing number of non-supersymmetric examples have been found~\cite{Bah:2020ogh,Bah:2021rki,Bah:2022yji,Bah:2023ows,Dulac:2024cso,Chakraborty:2025ger,Ganchev:2021pgs,Ganchev:2023sth,Heidmann:2023kry,Houppe:2024hyj,Dima:2025tjz,Bianchi:2025uis,Heidmann:2025pbb} (see~\cite{Bena:2025pcy} for a review). All of these geometries represent gravitational solitons realized by non-trivial topology, induced by the deformation of extra compact dimensions, and electromagnetic fluxes~\cite{Gibbons:2013tqa,deLange:2015gca}.

Topological solitons, as we will refer to the aforementioned objects in this work, are dramatically different from BHs near their surface. The latter feature an event horizon, which for an infalling observer is nothing but pure vacuum, yet it marks the point of no return for any such probe -- rendering the inner boundary of a BH spacetime perfectly absorbing. Topological solitons, on the other hand, replace the horizon with a smooth end of spacetime -- a ``cap'', which, furthermore, supports non-negligible horizon-scale microstructure. This results in an inner boundary that is very much the opposite of the BHs of GR: classically the cap is perfectly reflecting, and its physics are certainly not of an empty spacetime. Such a dramatic change should surely be expressed in certain properties that topological solitons have or lack in contrast to BHs. Our aim here is to elucidate one such feature.

However, given these stark differences, how is it that topological solitons provide a good fit to the criteria for BH mimickers exhibited earlier? Most of the GW observations up to date are a consequence of the physics at the unstable photon sphere in the exterior of the spacetime, which effectively governs the linear response to perturbations at early times, irrespective of the presence or absence of a horizon~\cite{Cardoso:2016rao,Cardoso:2017cqb}. The topological solitons we are interested in all possess such a photon sphere, and, in specific cases, it has indeed been shown that the prompt response as well as the quasinormal modes (QNMs) relative to neutral, massless (or massive but radiative) fields exhibit a spectrum that parallels the one of BHs, although other sectors with new properties also emerge~\cite{Heidmann:2023ojf,Bianchi:2023sfs,Dima:2024cok,Dima:2025zot}. In addition, particle motion as well as gravitational and electromagnetic fluctuations have been explored in~\cite{Dima:2024cok,Bena:2024hoh,Dima:2025zot,Melis:2025iaw,Bini:2025ltr,Bianchi:2025aei,Bini:2025qyn,DiRusso:2025lip,Bianchi:2024rod}, giving strong evidence for their linear stability and again demonstrating important similarities with BHs. Finally, with respect to their tidal deformability, topological solitons are strikingly similar to magnetic BHs, both exhibiting vanishing tidal Love numbers relative to neutral fields, but non-vanishing ones when charged probes are considered, as shown recently in Ref.~\cite{Pereniguez:2025jxq}.

While some of these studies consider solitons that cannot be made arbitrarily compact independently of their charge, the construction of realistic, neutral topological solitons~\cite{Bah:2022yji,Bah:2023ows,Heidmann:2023kry,Chakraborty:2025ger} is indeed possible. They also display a variety of properties, including the apparent size of their shadow and the scattering of null geodesics~\cite{Heidmann:2022ehn}, which are remarkably similar to those of their BH counterparts.

With this in mind, we want to approach the question of astrophysical significance of topological solitons from another direction. If they indeed provide a viable BH alternative stemming from a well-defined theory, yet possess features that are in stark contrast with those of BHs -- specifically their smooth inner boundary and the presence of microstructure there -- then we should equally try to characterize the properties that clearly distinguish them from the BHs of GR.

One such aspect that has remained largely unexplored concerns the bound states of massive fields around compact objects. In the case of BHs, it is well understood that such fields -- for example, dark matter candidates -- can grow spontaneously, forming clouds that are bound to the hole, a system known as the gravitational atom~\cite{Damour:1976kh,Zouros:1979iw,Detweiler:1980uk}. The mechanism yielding these clouds is the superradiant amplification due to the hole's rotation, so, by definition, such gravitational atoms are at most quasi-stationary states, and require that the central hole rotates (see~\cite{Brito:2015oca} for a review). These systems have been widely studied in the literature and have been considered to test possible dark matter candidates~\cite{Arvanitaki:2009fg,Arvanitaki:2010sy}, to obtain photon mass bounds~\cite{Pani:2012vp,Baumann:2019eav}, and have been proposed as new sources of gravitational~\cite{Yoshino:2013ofa,Brito:2014wla,Arvanitaki:2014wva} and electromagnetic radiation~\cite{Rosa:2017ury,Ikeda:2018nhb,Chen:2019fsq}. Moreover, when studying binaries, the cloud's presence can affect the orbital dynamics to varying extents, depending on the cloud's self-gravity~\cite{Ferreira:2017pth,Bar:2019pnz,GRAVITY:2019tuf,Vicente:2022ivh,Duque:2023seg,Dyson:2025dlj,Tomaselli:2024dbw,Tomaselli:2024bdd,Tomaselli:2023ysb}, and tidal effects can further enrich the binary dynamics~\cite{Baumann:2018vus,Berti:2019wnn,Cardoso:2020hca,Zhang:2018kib,Baumann:2019ztm}. 

In the case of topological solitons, the perfectly reflecting inner boundary condition, resulting from the smooth end of spacetime in their interior, would imply that the states of such massive fields are not only bound at infinity, but also experience no dissipation at the surface of the soliton. We expect then that topological solitons will display a normal mode spectrum in addition to their QNMs -- a property that is unattainable for the BHs of GR. This clear difference between BHs and TSs in the response to massive perturbations undoubtedly opens an interesting new avenue for exploration in view of the vast amount of research that has followed from the discovery of superradiant clouds around BHs.

In this work we confirm this expectation in the case of one of the simplest such geometries that is also one of the most widely studied ones: \textit{topological stars} (TSs)~\cite{Bah:2020ogh,Bah:2020pdz}. These are a one-parameter family of static solutions to the Einstein--Maxwell theory in five spacetime dimensions. Far from the star, the geometry is that of a four-dimensional Minkowski space endowed with a compact, circular extra dimension, as in the traditional Kaluza--Klein set up. Closer to the star, the geometry is smooth and supported by electromagnetic fluxes. Despite their simplicity, TSs contain the fundamental ingredients of any microstate geometry: a horizonless, smooth end-of-space, supported by electromagnetic flux. We probe them with an electrically charged, massive scalar field minimally coupled to the Einstein--Maxwell theory. We provide a comprehensive characterization of its bound states, that is, the states that are exponentially suppressed at infinity. Among our results, we show that any such mode is strictly stationary, thus proving that TSs are linearly stable under such perturbations with a characteristic normal mode spectrum. In this sense, TSs yield a more genuine notion of a gravitational atom than BHs, since the latter are only quasi-stationary and, in fact, unstable. We show that, in certain regimes, TS gravitational atoms share similarities with their BH counterparts, but they are, in general, markedly different. In addition, the KK momentum and electric charge endow the space of states with a rich structure, mostly absent in the BH case, which we analyze in detail. At the end of the paper, we discuss possible formation mechanisms of TS gravitational atoms focusing on potential instabilities of different nature.

While our work focuses on the simple TS, we believe that our results on the existence of normal modes should be widely applicable to any horizonless geometry with perfectly reflecting inner boundary conditions. Irrespective of whether generalizations are possible, we believe that simple topological solitons are excellent, UV-motivated toy models for exploring horizonless physics with the aim of learning valuable lessons for the future. 
Building a smooth, horizonless geometry that mimics Kerr BHs as closely as possible in our observations is currently a hard technical challenge, but no solid theoretical arguments to the contrary are known~\cite{Chakraborty:2025ger}. In light of this peak into the future, we would like to emphasize another property of topological solitons that is in definite contrast to BHs and requires more attention. As we clarified at the beginning of this article, classically, the end product of gravitational collapse for sufficiently dense energy distributions is always a BH. The dynamical formation of topological solitons is thus necessarily a process that requires physics beyond classical GR. 

This paper is structured as follows. First, in Section~\ref{sec:Summary} we summarize our results for the reader's convenience. Next, in Section~\ref{sec:MassiveBSGeneral} we discuss generic aspects of massive bound states of TSs. This includes proving that bound modes are strictly normal and showing that in the long (short) Compton wavelength regime, modes are hydrogenic (governed by timelike geodesics). In Section~\ref{Sec:NormalSpectrumTS} we provide a thorough analysis of TS normal modes, discussing their regime of existence, classifying them, and also computing their spectrum numerically, including the scenarios where the field carries KK momentum and is electrically charged. We conclude in Section~\ref{sec:discussion} with a discussion of our results and open questions.

\subsection{Executive summary}\label{sec:Summary}

Our results fall into two classes: those that are expected to hold in a qualitatively similar way for compact objects other than TS, and those that are TS-specific. We summarize them here. We remind the reader that TSs are characterized by two length scales, the TS's size $r_{b}$ and the length of the compact extra-dimension $R_{y}$.
\subsubsection{Generic results}

\paragraph{\textbf{Bound modes are strictly normal:}} 
massive modes can be exponentially suppressed at infinity if their mass $\mu$ is sufficiently large to endow them with a negative ``binding energy''. We show that, in those cases, the modes must be strictly normal, that is, their frequencies are exactly real, $\omega\in\mathbb{R}$. We show this geometrically, employing conservation laws, and our argument does not rely on the (spatial) separability properties of the solutions. A similar statement is expected to hold for any compact object supported by geometry and topology where, in stark contrast with BHs, the regularity of fluctuations at the spacetime's center entails absence of dissipation.

\begin{figure}[h]
	\includegraphics[width=0.85\linewidth]{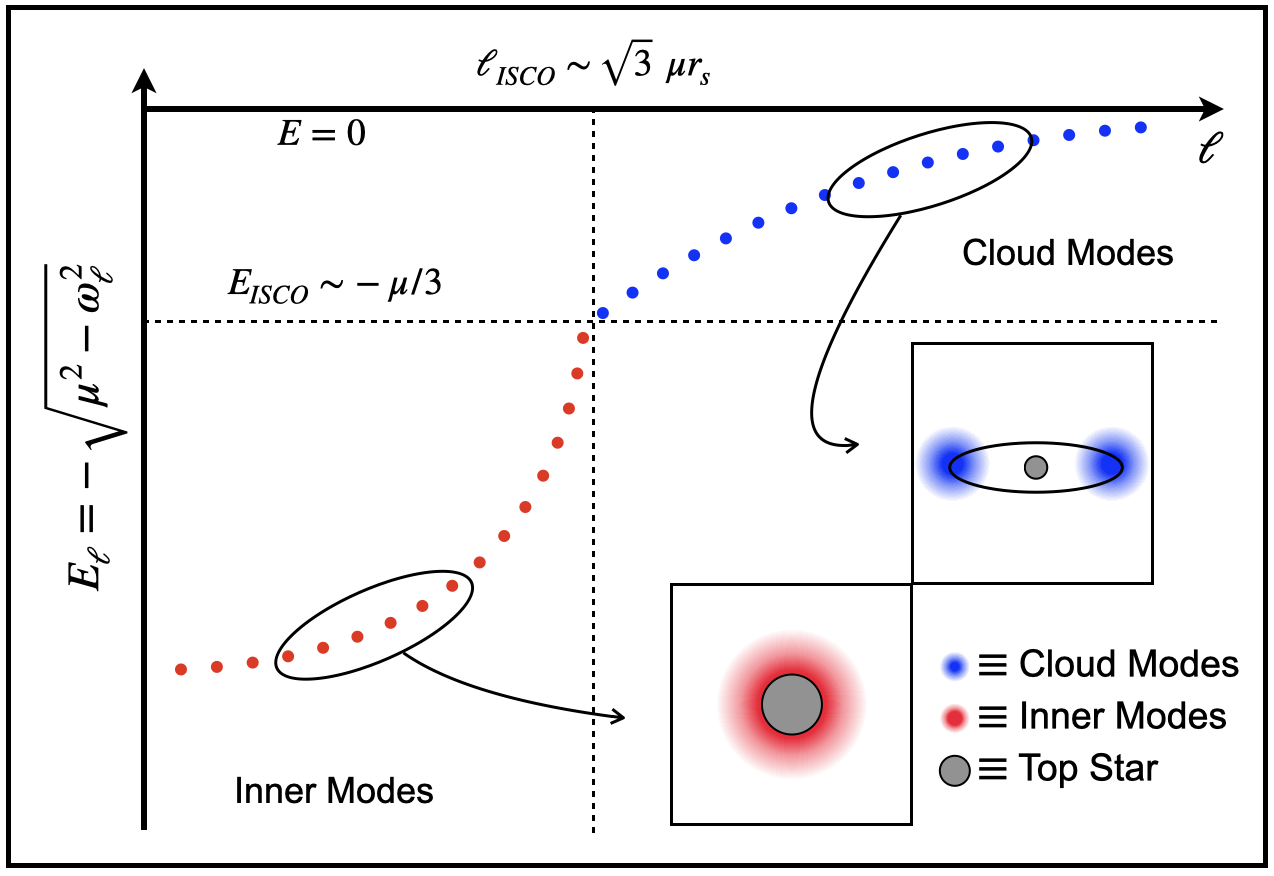} 
	\caption{Schematic energy spectrum of normal modes at fixed overtone, $E_{\ell}=-\sqrt{\mu^{2}-\omega_{\ell}^{2}}$, as a function of the harmonic multipole $\ell$. Red (blue) dots correspond to inner (cloud) modes. Cloud modes are associated to stable circular geodesics, and terminate approximately at the mode associated to the ISCO, indicated with dashed thin lines. Inner modes instead fluctuate close to the star and are sensitive to its structure.}
	\label{Fig:Intro1} 
\end{figure}

\paragraph{\textbf{Geometric particle limit and cloud versus inner modes:}} considering neutral probes (no charge or KK momentum), in the limit $\mu r_{b}\gg1$, we show that solutions are governed by congruences of timelike geodesics. This allows us to identify two broad families of modes, dubbed \textit{cloud} and \textit{inner} modes. The former have a large angular momentum $\ell \gg \mu r_{b}$ and are exponentially localized around stable circular geodesics. They cease to exist for angular momenta lower than $\ell \sim \sqrt{3} r_{s}\mu$, where $r_{s}$ is a length fixed by $r_{b}$ and $R_{y}$
and related to the ISCO by $r_{\rm ISCO}=3 r_{s}$. The behavior of modes changes abruptly below this threshold and becomes dictated by the inner structure of the geometry: these are the inner modes. The separation between the two families, at $\ell \sim \sqrt{3} r_{s}\mu$, is associated with the ISCO, and the modes inherit some of its properties. This is illustrated in Fig.~\ref{Fig:Intro1}. 

\paragraph{\textbf{The hydrogenic limit:}} in the opposite regime, where $\mu r_{b}\ll1$, the normal mode spectrum (restricting attention to zero-KK modes) becomes hydrogenic,
\begin{equation}
    \omega_{n}\approx \mu\left(1-\frac{\mu^{2}r_{s}^{2}}{8}\frac{1}{n_{\text{H}}^{2}}\right)\, , \quad  \left(n_{\text{H}}=1,2,... \, , \ \ n_{\text{H}}>\ell\right).
\end{equation}
A similar behavior is also present in the superradiant quasi-bound states of rotating BHs~\cite{DetweilerGA,Baumann:2019eav}, in the limit $\mu M\ll1$, where $M$ is the BH's mass. One difference is the fine structure constant, which here is proportional to $\mu r_{s}$, while it is proportional to $\sim\mu M$ for BHs. However, the most important difference is that in the TS case these modes are truly stationary states, unlike for BHs, where they are generically unstable~\cite{Hod:2012px,Herdeiro:2014goa,Carretero:2025eqh,Nicoules:2025bhh}.

\subsubsection{TS-specific results}

\paragraph{\textbf{TS and mixed modes:}} for probes with no KK momentum (yet possibly charged), the gravitational potential of TSs for bound modes can yield either one or two classically allowed regions where modes can be localized. This leads to a classification of bound states in TS modes, cloud modes, or mixed modes. The former two have a single classically allowed region, with TS modes localized within the ISCO while cloud modes are located outside of it. Mixed modes have two classically allowed regions separated by a centrifugal barrier. In the limit $\mu r_{b}\gg1$, the cloud modes are dictated by the behavior of stable circular geodesics, as discussed above. The addition of KK momentum enriches the mode structure by introducing additional potential barriers, as discussed next.

\paragraph{\textbf{KK ionization and potential barriers:}} normal modes of TSs can carry KK momentum, $k=p/R_{y}$, where $p$ is an integer and $R_{y}$ the radius of the extra dimension. Far from the star, this yields the usual tower of mass corrections to the mode. Closer to the star, though, KK momentum results in a potential barrier that prevents the mode from reaching the star's center. In fact, for large enough momentum modes cannot remain bound to the star and are radiated to infinity. More precisely, we find that a KK mode can be bound only if (taking $p>0$ without loss of generality)
\begin{equation}\label{eq:KKion}
    p<\frac{R_{y}/\lambda}{\sqrt{\frac{r_{b}}{r_{s}}-1}}\equiv p_{\star}\,,
\end{equation}
where $\lambda=1/\mu$ is the field's Compton wavelength. Modes with $p\geq p_{\star}$ are necessarily radiated to infinity, so $p_{\star}$ defines the KK ionization threshold. Assuming that Eq.~\eqref{eq:KKion} holds, we find that, similarly to the zero-KK case, in the regime $\mu r_{b}\ll1$ the spectrum becomes hydrogenic,
\begin{equation}
\label{eq:hydrogenic_spectrum}    \omega_{n_{\text{H}}}\approx\sqrt{\mu^{2}+k^{2}}\left[1-\frac{\left(\mu^{2}-(r_{b}/r_{s}-1)k^{2}\right)^{2}r_{s}^{2}}{4\left(\mu^{2}+k^{2}\right)}\frac{1}{2n_{H}^{2}}\right]\,,
\end{equation}
where $n_{\text{H}}=1,2,...$ and $n_{\text{H}}>\ell$.

\begin{figure}[h]
	\includegraphics[width=0.95\linewidth]{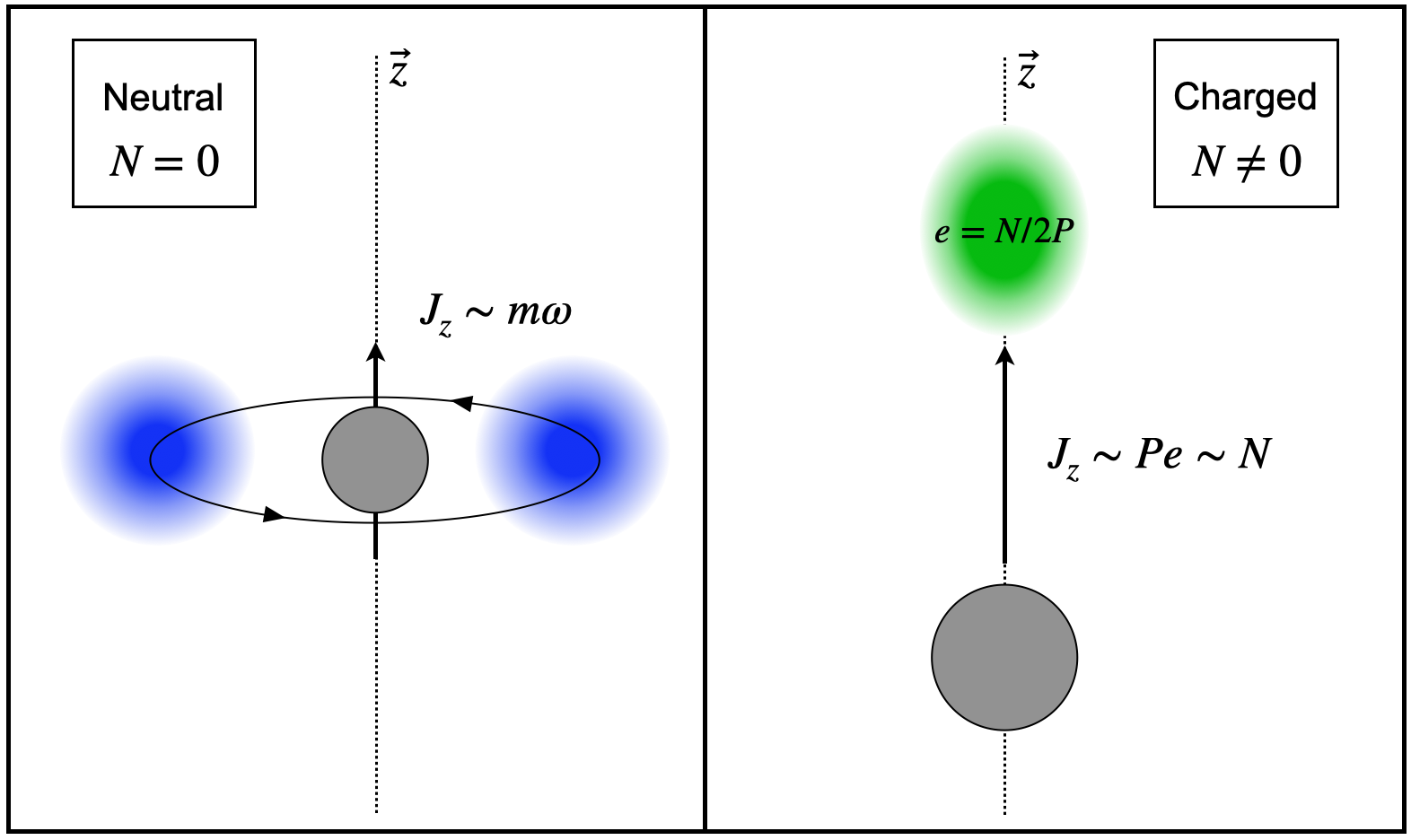} 
	\caption{Schematic representation of the spatial distribution of neutral (left panel) and charged (right panel) modes. Both states carry angular momentum in the $z$ direction, in the neutral case due to orbital motion, and in the charged one by the Thomson dipole effect.}
	\label{Fig:Intro2} 
    \end{figure}

\paragraph{\textbf{Charged modes yield gravitating Thomson dipoles:}} if the scalar field is electrically charged, it feels the magnetic monopole carried by the star, and its fundamental charge is quantized according to Dirac's condition~\cite{Dirac:1931kp}
\begin{equation}
    e=N/2P\,,\quad N=0,\pm1,\pm2,...
\end{equation}
where $P$ is the TS's monopole charge. While the classification of neutral fields, based on their radial distribution, extends similarly to charged modes, the angular distribution of the latter changes significantly. In particular, charged modes can become localized at the north or south axis of the star, forming an axially symmetric configuration, see Fig.~\ref{Fig:Intro2}. We show that these states carry more angular momentum than their neutral counterparts, when normalized to have the same energy. This yields a gravitating version of a Thomson dipole in the strong gravity regime (recall that a Thomson dipole is the system formed by an electric and a magnetic charge~\cite{thomson_2009}). It constitutes an alternative to previous constructions where one attempts to form a gravitating Thomson dipole with BHs and point particles~\cite{Garfinkle:1990zx,Bunster:2007sn,Kim:2007ca,Dyson:2023ujk}, and it might possess a nonlinear completion similar to that of the superradiant threshold modes of magnetic BHs~\cite{Pereniguez:2024fkn,Cunha:2024gke}. 

\section{Massive bound states of Topological Stars}\label{sec:MassiveBSGeneral}

A massive, charged scalar field minimally coupled to the Einstein--Maxwell theory in five spacetime dimensions is governed by the action ($\kappa_{5}=\hbar=c=1$)
	\begin{equation}\label{eq:theory}
		S=\int d^{5}\sqrt{-g}\left\{\frac{1}{2}R-\frac{1}{4}F^{2}-\frac{1}{2}\left(D_{\mu}\bar{\Phi}D^{\mu}\Phi+\mu^{2}\bar{\Phi}\Phi\right)\right\}\,,
	\end{equation}
where $D_{\mu}=\nabla_{\mu}+i e A_{\mu}$ is the gauge-covariant derivative, and the $U(1)$-action is
	\begin{equation}\label{eq:gauge}
		\Phi\mapsto e^{i \alpha(x)}\Phi\, ,\quad A_{\mu}\mapsto A_{\mu}-\frac{1}{e}\nabla_{\mu}\alpha(x)\,,
	\end{equation}
parametrized by a local real function $\alpha(x)$. Here, $\mu$ is the field mass and $e$ its Maxwell coupling constant, or fundamental charge. The TS-black string solution~\cite{Bah:2020ogh} is a 2-parameter family with vanishing scalar, $\Phi=0$, given by
\begin{align}\label{eq:TopStar}\notag
		ds^{2}&=-f_{s}(r)dt^{2}+f_{b}(r)dy^{2}+\frac{dr^{2}}{f_{s}(r)f_{b}(r)}+r^{2}d\Omega^{2}_{2}\,,\\
		A&=-P\,\cos{\theta}\,d\phi\, ,\quad f_{s,b}=1-\frac{r_{s,b}}{r}\, ,
\end{align}
where $r_{s,b}$ are free parameters and the magnetic charge $P$ is fixed to
\begin{equation}
    P=\pm\,\left(3\,r_{s}\,r_{b}/2 \right)^{1/2}\, .
\end{equation}
It describes a magnetically charged black string if $r_{s}>r_{b}>0$, or a TS if $r_{b}>r_{s}>0$. Our prime interest is in the TS regime, so we restrict to that henceforth. In that case, to make sense of the geometry at $r=r_{b}$, the extra dimension $y$ should be identified as $y\sim 2\pi R_{y}$, and the period $R_{y}$ is fixed by the orbifold condition
\begin{equation}\label{eq:orbifold}
r_{s}=r_{b}\left(1-4\frac{r_{b}^{2}}{K^{2} R_{y}^{2}}\right)\, , \quad (K\in \mathbb{N})\, ,
\end{equation}
so $K$ labels one-parameter families of solutions. The family with $K=1$ is smooth at $r=r_{b}$ and the global space topology is $R^{2}\times S^{2}$, while $K>1$ describes an orbifold geometry $R^{2}/\mathbb{Z}_{K}\times S^{2}$. Fluctuations about those backgrounds include metric ($\delta g_{\mu\nu}$), electromagnetic ($\delta A_{\mu}$) and scalar ($\delta\Phi$) perturbations. However, due to the vanishing scalar background, first-order scalar fluctuations decouple from the gravitational and electromagnetic ones, and satisfy (henceforth we write $\delta\Phi\equiv
\Phi$ to simplify notation),
\begin{equation}\label{eq:waveeq}
    \left(D^{\mu}D_{\mu}-\mu^{2}\right)\Phi=0\, .
\end{equation}
Our aim in this section is to establish some general properties of solutions of Eq.~\eqref{eq:waveeq}. This is similar to the case of neutral fields, previously considered~\cite{Heidmann:2023ojf,Bianchi:2023sfs}, except for the angular analysis, which presents substantial differences. First, to ensure that solutions of Eq.~\eqref{eq:waveeq} belong to a smooth $U(1)$-bundle, one needs to impose Dirac's quantization condition~\cite{Dirac:1931kp},
\begin{equation}\label{eq:DiracQuant}
	2Pe= N\,, \ \ \ N\in Z\, .
\end{equation}
Separable solutions of Eq.~\eqref{eq:waveeq} can be constructed by employing the Wu-Yang monopole harmonics $Y_{N\ell m}(\theta,\phi)=\mathcal{P}_{N,\ell, m}(\theta)e^{i m\phi}$~\cite{Wu:1976ge} (we review them in Appendix \ref{A1}), 
\begin{equation}\label{eq:mode}
\begin{aligned}
	\Phi&=e^{-i\,\omega\,t+i\,k\,y+i\,m\,\phi}\psi(r)\mathcal{P}_{N,\ell, m}(\theta)\, ,
\end{aligned}
\end{equation}
where the KK momentum is quantized as (without loss of generality, we take $p$ as a non-negative integer)
\begin{equation}
	k=p/ R_{y}\, ,
\end{equation}
and 
\begin{equation}
	\ell=\frac{\lvert N\rvert}{2},\frac{\lvert N\rvert}{2}+1,...\, ,\quad m=-\ell, -\ell+1,...,  \ell\, .
\end{equation}
With that, the radial function is subject to
\begin{widetext}
    \begin{equation}\label{eq:radeq}
	(r-r_{S})(r-r_{B})\psi''(r)+(2r-r_{B}-r_{S})\psi'(r)+\left(\frac{r^{3}\left(\omega ^2-\mu ^2\right)+\mu ^2  r_{S}r^2}{r-r_{S}}-\frac{k^2 r^3}{r-r_{B}}-\Lambda\right)\psi(r)=0\, ,
\end{equation}
\end{widetext}
where 
\begin{equation}
    \Lambda=\ell(\ell+1)-(N/2)^{2}\geq0\, .
\end{equation}
The behavior of the radial function at $r=r_b$ is fixed by regularity requirements,
\begin{equation}\label{eq:rb}
	\begin{aligned}
		\psi(r)&\sim (r-r_b)^{p\,K/2} \quad \left(r\sim r_{b}\right)\, .
		\end{aligned} 
\end{equation}
Focusing on the case of a smooth background geometry, with $K=1$, this regularity condition implies\footnote{This can be seen by replacing $(r,y)$ with coordinates $(x,z)$ that are well-defined close to $r=r_{b}$, given by $ r=r_{b}+\left(x^{2}+z^{2}\right)/L$ and $y=\sqrt{L r_{b}}\arctan{\left(z/x\right)}$, with $L\equiv\frac{4 r_{b}^{2}}{r_{b}-r_{s}}$, and showing that $ \frac{\partial^{m+l}}{\partial x ^{m}\partial z^{l}}\varphi\sim e^{i\frac{p-(m+l)}{ R_{y}}y}\left(r-r_{B}\right)^{\frac{p-(m+l)}{2}}$.} that the differentiability class of $\Phi$ in the neighborhood of $r=r_{b}$ is set by the units of KK momentum,
\begin{equation}\label{eq:Dif}
	\Phi\in \begin{cases}C^{p}\left(M\right) \quad &\text{if} \quad p>0 \\
	C^{\infty}\left(M\right) \quad &\text{if} \quad p=0 \end{cases}
\end{equation}
where $M$ denotes the spacetime manifold. This shows that only the zero-KK modes are smooth functions, a fact that can be interpreted as the scalar field probing a delta-source at the star's center. However, as we will see below, the conserved physical currents associated to $\Phi$ (e.g., the energy) are first-derivative functionals of $\Phi$, so the fact that in all cases the differentiability class of $\Phi$ is $C^{p\geq 1}(M)$ guarantees that Eq.~\eqref{eq:rb} indeed corresponds to physically regular solutions. 

At infinity, solutions can present two linearly independent behaviors, $\psi(r)\sim e^{\pm i \sqrt{\omega^{2}-k^{2}-\mu^{2}} \ r}$, and requiring that only one of them is present,
\begin{equation}\label{eq:rinf}
	\begin{aligned}
		\psi(r)&\sim e^{i \sqrt{\omega^{2}-k^{2}-\mu^{2}}\  r} \quad \left(r\to \infty\right)\, ,
		\end{aligned} 
\end{equation}
defines a characteristic value problem for the frequencies, $\omega$, which in general will allow only a discrete set of solutions. Assuming $\omega$ satisfies such problem, the solutions fall into two classes:
\begin{equation}\label{eq:QNMvsBS}
    \begin{aligned}
        \text{QNMs:}&\quad \text{Im}\left[\sqrt{\omega^{2}-k^{2}-\mu^{2}}\right]<0\,,\\
        \text{Bound states:}&\quad \text{Im}\left[\sqrt{\omega^{2}-k^{2}-\mu^{2}}\right]>0\,.
    \end{aligned}
\end{equation}
QNMs are solutions that propagate to infinity, oscillating and decaying at characteristic frequencies and damping times. These solutions have been intensively studied in the literature and compared to the QNMs of BHs (see e.g.~\cite{Heidmann:2023ojf,Bianchi:2023sfs}). In this work, we focus on bound states that, unlike QNMs, are exponentially suppressed at infinity. 

In general, in BH spacetimes bound states decay fast in time, being absorbed by the hole. However, in the case of rotating BHs bound states can grow in time due to superradiant instabilities~\cite{Brito:2015oca}. At the threshold separating time-growing and time-decaying solutions lie the stationary states, which are bound to the hole and oscillate at a specific frequency (the superradiant frequency), but do not decay in time~\cite{Hod:2012px,Herdeiro:2014goa}. This yields the notion of a gravitational atom, discussed in the introduction. 

In the case of TSs, bound states are not only unable to leak out at infinity, but also lack a dissipation mechanism at the star surface, due to the event horizon being replaced by a perfectly reflecting boundary. Indeed, in the next section we show that bound states in TS spacetimes are strictly normal modes, that is, they oscillate in time, but do not decay. These give rise to a genuine realization of a gravitational atom, superseding systems involving BHs in the sense that they are stable configurations -- the only dissipation mechanism is through the emission of gravitational and electromagnetic waves. In addition, these states turn out to exhibit a rich structure, that we discuss at length in Section~\ref{Sec:NormalSpectrumTS}. 

Before that, here we establish some generic facts. Namely, we first prove that any bound state of a TS must be strictly a normal mode. Next, we show that in the high-frequency regime, where the field's Compton wavelength $\lambda=1/\mu$ is short compared to the TS's size, $\mu r_{b}\gg1$, normal modes are described by congruences of timelike geodesics. This allows us to identify two broad classes of modes, dubbed cloud modes and inner modes, with qualitatively distinct properties. Finally, we show that in the opposite regime, where $\mu r_{b}\ll1$, the spectrum becomes hydrogenic. This is similar to the states of the BH gravitational atom~\cite{DetweilerGA,Baumann:2019eav} but, in our case, the clouds are stable and non-decaying due to the absence of an event horizon.

\subsection{Proving that bound modes are normal modes}\label{ssec:normalProof}

Heuristically, bound modes of TSs should persist over time due to energy conservation, as being exponentially suppressed at infinity and in the absence of horizons there is no available channel for their dissipation. An elegant way to prove this is to employ the conservation laws associated to fields propagating on the background \eqref{eq:TopStar}, as we do next. Alternatively, one could simply refer to the hermiticity properties of the ODE problem satisfied by $R(r)$ in Eq.~\eqref{eq:mode}, in the case of bound states. We prefer the former approach, since we find it more geometric and physically insightful and it does not rely on unphysical statements, such as the separability properties of the spatial coordinates. 

To construct the conserved currents, we notice that every Killing vector $X$ of the background \eqref{eq:TopStar} has an associated \textit{momentum map} $\mathcal{P}_{X}$, a function satisfying 
	\begin{equation}
		\nabla_{\mu}\mathcal{P}_{X}+X^{\nu}F_{\nu\mu}=0\, .
	\end{equation}
Then, if $\Phi$ satisfies its equation of motion, the following local functional\footnote{Here, $\boldsymbol{\epsilon}$ is the volume form and the notation $\boldsymbol{\epsilon}_{\mu}T^{\mu}$ stands for $\boldsymbol{\epsilon}_{\mu \mu_{1}...\mu_{d-1}}T^{\mu}$, in contrast with our conventions for the hodge dual, $\star T_{\mu_{1}...\mu_{d-1}}=\boldsymbol{\epsilon}_{\mu_{1}...\mu_{d-1}\nu}T^{\nu}$. To avoid cluttering notation, we do not write the indices of differential forms, and just use bold symbols to indicate that omission, following the notational conventions in~\cite{Wald:1993nt}.} is conserved:
\begin{equation}\label{eq:ConForm}
		\mathbf{I}_{X}\left[\Phi\right]\coloneqq \left\{e \mathcal{P}_{X} J^{\Phi}_{\nu}+X^{\mu}T^{\Phi}_{\mu\nu}\right\}\boldsymbol{\epsilon}^{\nu}\, , \quad d\mathbf{I}_{X}[\Phi]=0\, ,
\end{equation}
where $T^{\Phi}_{\mu\nu}$ and $J^{\Phi}_{\mu}$ are the field's energy-momentum tensor and current density:

\begin{equation}\label{eq:emTens}
	\begin{aligned}
			T^{\Phi}_{\mu\nu}&=D_{(\mu}\bar{\Phi}D_{\nu)}\Phi-\frac{1}{2}\left(D_{\alpha}\bar{\Phi}  D^{\alpha}\Phi+\mu^{2} \bar{\Phi}\Phi\right)g_{\mu\nu}\,, \\
            J^{\Phi}_{\mu}&=\frac{i}{2}\left(\bar{\Phi}D_{\mu}\Phi-\Phi D_{\mu}\bar{\Phi}\right)\, .
	\end{aligned} 
\end{equation}

\

\

\begin{figure}[h]
	\centering
	\includegraphics[width=0.75\linewidth]{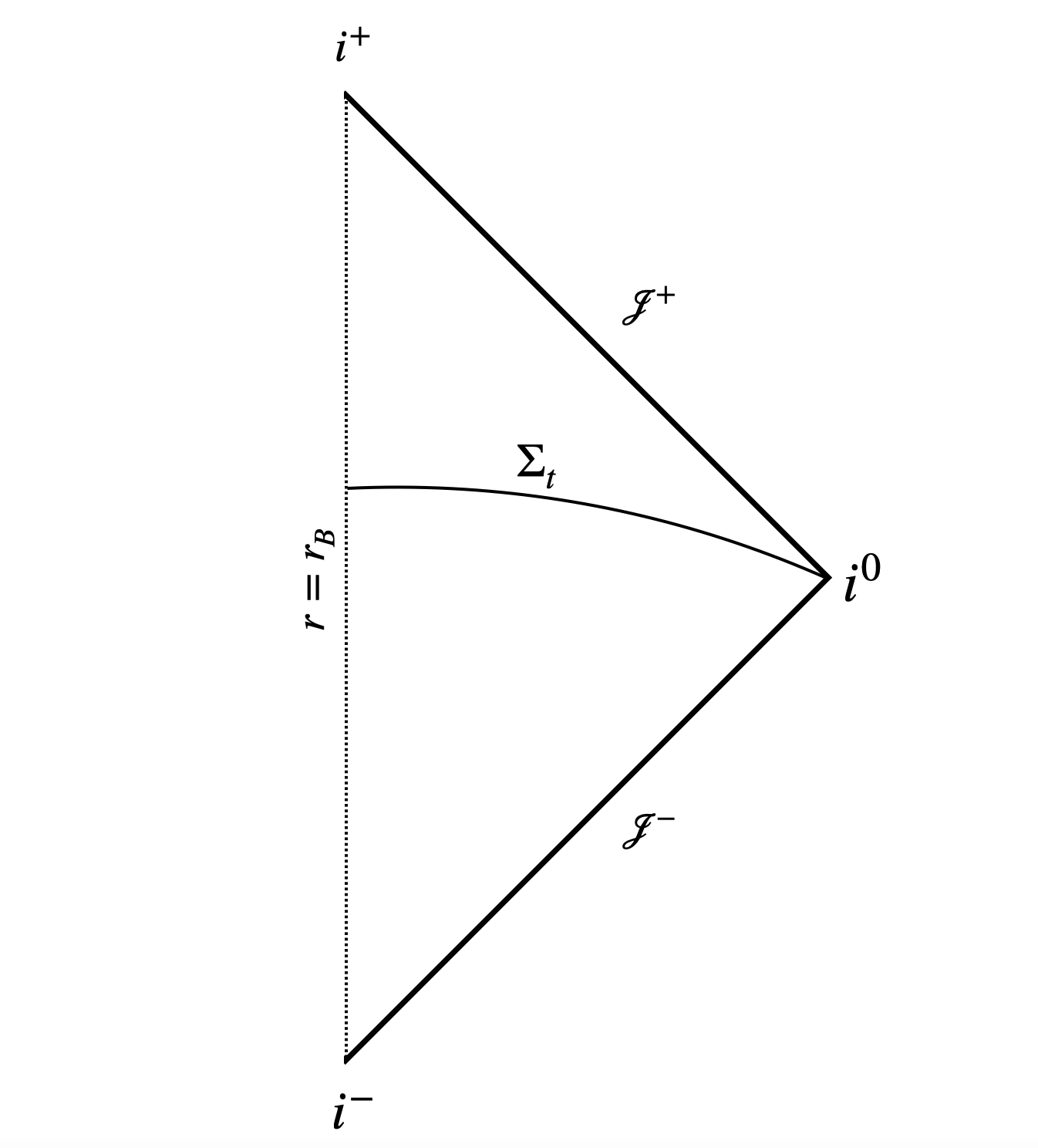} 
	\caption{Surface $\Sigma_{t}$ in a Carter--Penrose diagram of a TS. The diagram represents the quotient by the $y$-circle and the 2-sphere.}
	\label{Fig:1} 
\end{figure}

The conservation equation~\eqref{eq:ConForm} on-shell follows by construction from the gauge symmetries of the theory (see~\cite{Elgood:2020svt,Ortin:2022uxa,Dyson:2023ujk}), but can also be verified explicitly using the Noether identities given in Appendix \ref{A2}. For the TS \eqref{eq:TopStar}, the Killing vectors $\partial_{t},\partial_{\phi}$ and $\partial_{y}$ have associated the momentum maps,
	\begin{equation}
		\mathcal{P}_{\partial_{t}}=\mathcal{P}_{\partial_{y}}=0\,, \quad \mathcal{P}_{\partial_{\phi}}=-P \cos\theta\, ,
	\end{equation}
	and the associated conserved functionals can be interpreted as changes in ADM mass, angular- and KK-momentum due to the presence of the scalar field~\cite{Ortin:2022uxa,Dyson:2023ujk,Ballesteros:2023iqb,Gomez-Fayren:2023wxk}. It will suffice to consider the energy functional associated to $\partial_{t}$. We shall evaluate it on a mode of the form \eqref{eq:mode}, assuming that the frequency possesses both real and imaginary parts,
	\begin{equation}
		\omega=\omega_{R}+i\omega_{I}\, .
	\end{equation}
    Then, integrating it on a constant time slice of \eqref{eq:TopStar}, denoted $\Sigma_{t}$ (see Fig.~\ref{Fig:1}), gives
    \begin{widetext}
	\begin{equation}
		\begin{aligned}\label{eq:Integral}
			E[\Phi]\coloneqq-\int_{\Sigma_{t}}\bold{I}_{\partial_{t}}\left[\Phi\right]=\pi R_{y}e^{2\omega_{I}t}\int_{r_{B}}^{\infty}dr\left\{\left(\frac{p^{2}}{R_{y}^{2}}\frac{r^{2}}{f_{B}(r)}+\lvert\omega\rvert^{2}\frac{r^{2}}{f_{S}(r)}+\mu^{2}r^{2}+\Lambda\right)\lvert\psi\rvert^{2}+r^{2}f_{B}(r)f_{S}(r)\lvert\psi'\rvert^{2}\right\}\, ,
		\end{aligned}
	\end{equation}
    \end{widetext}
	 where we assumed the orbifold condition \eqref{eq:orbifold} with $K=1$ for simplicity, i.e., the smooth TS, but the argument can be extended to the general background with $K\geq1$.
The fact that the modes are at least $C^{1}(M)$ in the vicinity of $r=r_{B}$, together with the bound-state condition in Eq.~\eqref{eq:QNMvsBS}, guarantee that the integral is finite. Then conservation of $E[\Phi]$ implies that it cannot depend on the time slice chosen,\footnote{One subtlety is that, for Stokes' theorem to apply in this case, modes should be at least $C^{2}(M)$ functions, which is not true for modes carrying precisely one unit of KK momentum, see Eq.~\eqref{eq:Dif}. In that case, one can modify the integration domain adding a timelike boundary at $r=r_{B}+\delta$, apply Stokes theorem, and show easily that the additional integral vanishes as $\delta\to0$. The same strategy can be used to generalize our argument to the TS with non-trivial orbifolds $K>1$.} so it must be that
\begin{equation}
    \omega_{I}=0\, .
\end{equation}
We can thus conclude that bound modes are strictly normal modes. The numerical and analytical results of the following sections confirm this fact, as expected. A similar strategy can be used in the case of BHs, where the integrals at two
time slices would differ by an integral across the event horizon (see~\cite{Pereniguez:2024fkn} for the related case of a magnetic BH). From the latter, it follows immediately that a bound state that is not dissipating through the horizon necessarily oscillates at the superradiant frequency and does not decay in time.

\subsection{The geometric particle limit $\mu r_{b}\gg1$:\\ inner vs. cloud modes}\label{ssec:geoAnalysis}

A great deal of intuition about the structure of normal modes can be gained by studying a geometric particle limit. Similarly to the usual geometric optics limit, this focuses on fluctuations whose characteristic wavelength is much shorter than the background's lengthscale. For simplicity, we will restrict ourselves to neutral modes, with no KK momentum. The idea consists in making a large-phase ansatz, assuming that the field's Compton wavelength $\lambda=1/\mu$ is much shorter than the star's size, i.e., $\lambda/r_{b}\ll1$. Writing
\begin{equation}\label{eq:geopartic}
	\Phi=\left[A(x)+ O(\lambda/r_{b})\right]e^{i\frac{S(x)}{\lambda/r_{b}}}\,, \
\end{equation}
inserting this into the wave equation and expanding in $\lambda/r_{b}$ one finds, to leading order,
\begin{equation}\label{eq:TimelikeGeod}
	\nabla^{\mu}S\nabla_{\mu}S=-1\,  .
\end{equation}
In terms of $k_{\mu}\equiv\nabla_{\mu}S$, this implies that 
\begin{equation}
	k^{\mu} k_{\mu}=-1\,, \quad k^{\mu}\nabla_{\mu}k_{\nu}=0\, ,
\end{equation}
so in the small-$\lambda$ limit the solutions to the wave equation are governed by congruences of timelike geodesics. To solve Eq.~\eqref{eq:TimelikeGeod} (restricting again to the case of zero KK momentum), we note that the principal function $S$ admits separable solutions of the form
\begin{equation}
		S=-\mathcal{E} t+r_{b}\mathcal{M} \phi+S_{R}(r)+ S_{\theta}(\theta)\,.
\end{equation}
Here, $\mathcal{E}$ and $\mathcal{M}$ are constants and $S_{R,\theta}$ satisfy the equations
\begin{equation}\label{eq:floweqs}
    S'_{R}(r)=\pm\sqrt{\mathcal{R}(r)}\, ,\quad S'_{\theta}(\theta)=\pm\sqrt{\Theta(\theta)}\, , 
\end{equation}
where 
\begin{equation}
	\begin{aligned}
            \mathcal{R}(r)&\equiv\frac{1}{f_{s}^{2}f_{b}}\left[\mathcal{E}^{2}-f_{s}\left(1+\frac{\mathcal{Q}+r_{b}^{2}\mathcal{M}^{2}}{r^{2}}\right)\right]\,,\\
		   \Theta(\theta)&\equiv \mathcal{Q}-r_{b}^{2}\mathcal{M}^{2}\cot^{2}\theta\, ,
	\end{aligned}
\end{equation}
with $\mathcal{Q}$ being a separation constant. With this, solutions at leading order in $\lambda$ are:
\begin{equation}\label{eq:LO}
	\Phi\sim e^{\frac{i}{\lambda}\left[-\mathcal{E} t+r_{b}\mathcal{M} \phi+S_{R}(r)+ S_{\theta}(\theta)\right]}\,,	
\end{equation}
so these correspond to mode solutions of the form \eqref{eq:mode} with frequency and angular number
\begin{equation}
    \omega=\mathcal{E}/\lambda\, ,\quad m=r_{b}\mathcal{M}/\lambda\, .
\end{equation}
So far, $\mathcal{E}$, $\mathcal{M}$ and $\mathcal{Q}$ are arbitrary (except for the condition that $r_{b}\mathcal{M}/\lambda\in Z$, to guarantee that the solution is single-valued). For reasons that will be clear soon, we shall fix them by requiring that $\mathcal{R}(r)$ and $\Theta(\theta)$ vanish and have maxima at some radius $r=r_{0}$ and at the equator $\theta=\pi/2$, respectively. That sets $\mathcal{Q}=0$, and fixes $\mathcal{E}$ and $r_{0}$ in terms of $\mathcal{M}$ as
\begin{equation}\label{eq:circOrbits}
\begin{aligned}
	r_{0}(\mathcal{M})&=r_{b}\frac{\mathcal{M}^{2}}{r_{s}/r_{b}}\left(1+\Delta\right)\, , \quad \Delta\equiv\sqrt{1-3(r_{s}/r_{b}\mathcal{M})^{2}}\,, \\
    \mathcal{E}^{2}(\mathcal{M})&=\frac{\left[(r_{s}/r_{b}\mathcal{M})^{2}-(1+\Delta)\right]^{2}}{\left(1+\Delta\right)\left(1-(3/2)(r_{s}/r_{b}\mathcal{M})^{2}+\Delta\right)}.
\end{aligned}
\end{equation}
With these choices, the solutions of Eq.~\eqref{eq:floweqs} for both $S_{\theta}(\theta)$ and $S_{R}(r)$ are purely imaginary, so the wave \eqref{eq:LO} is exponentially localized at the equator and at $r=r_{0}(\mathcal{M})$. At large values of $\mathcal{M}$, the frequency of the mode and the radial position at which it is localized are
\begin{equation}\label{eq:largeM}
	\omega=\frac{1}{\lambda}\left[1-\frac{(r_{s}/r_{b})^{2}}{8 }\frac{1}{\mathcal{M}^{2}}+...\right]\, ,\quad r_{0}=2 r_{b}\eta\mathcal{M}^{2}+...
\end{equation}
where the ellipsis denotes subleading terms in the large-$\mathcal{M}$ expansion. The modes are localized closer to the star as $\mathcal{M}$ decreases. However, below a threshold value $\mathcal{M}_{\rm min}=\sqrt{3} r_{s}/r_{b}$ the behavior changes qualitatively. For $\mathcal{M}=\mathcal{M}_{\rm min}$, one has
\begin{equation}\label{eq:ISCOenergy}
	\omega=\frac{1}{\lambda}\frac{2\sqrt{2}}{3}=\frac{1}{\lambda}\mathcal{E}_{\rm ISCO} \, , \quad r_{0}=3r_{s}\equiv r_{\rm ISCO}\, ,
\end{equation}
where we used that the ISCO of TSs lies at $3r_{s}$. However, the mode now is not localized at $r=r_{\rm ISCO}$, since that becomes an inflection point of $\mathcal{R}(r)$ rather than a maximum. We have $\mathcal{R}(r)>0$ for $r<r_{\rm ISCO}$ and $\mathcal{R}(r)<0$ for $r>r_{\rm ISCO}$, so the corresponding solution \eqref{eq:LO} oscillates within the region $r<r_{\rm ISCO}$, but is exponentially suppressed beyond it. 

This analysis predicts the existence of two broad classes of modes in the regime $\mu r_{b}\gg1$. For large angular momentum $m\gg\mu r_{b}$ (so $\ell\gg \mu r_{b}$), the solutions of the massive wave equation are described by Eq.~\eqref{eq:LO}, with $\mathcal{E}$ given by Eq.~\eqref{eq:largeM} and the modes localized far away from the star, at the radius predicted by Eq.~\eqref{eq:largeM}. We shall refer to these as \textit{cloud modes}. On the other hand, for lower angular momentum $\ell\ll\,r_{b}\mu$ modes can probe the star's center, so we refer to them as \textit{inner modes}. The boundary between cloud and inner modes occurs at $\ell\sim\sqrt{3} \mu r_{s}$, where the modes oscillate at approximately the ISCO energy, as predicted in Eq.~\eqref{eq:ISCOenergy}. We expect any horizonless compact object to admit a similar mode classification in the small-$\lambda$ regime. However, for finite values of $\lambda$, the inner modes will exhibit properties that depend on the object's structure. In the next section, we confirm our predictions at small $\lambda$, and establish a more accurate classification of the TS inner modes for arbitrary values of the parameters.

\subsection{The hydrogenic regime $\mu r_{b}\ll1$}

Let us introduce the dimensionless coordinate $z=\mu r$. In the regime $\mu r_{b}\ll1$ (and hence $\mu r_{s}\ll1$ since $r_{s}<r_{b}$), the range of $z$ is $\sim(0,\infty)$ and the radial equation \eqref{eq:radeq} to leading order in $\mu r_{s}$ becomes
\begin{widetext}
    \begin{equation}\label{eq:SchApprox}
    z^{2}\psi''(z)+2z\psi'(z) +\left[\left(\frac{\omega^{2}}{\mu^{2}}-\frac{k^{2}}{\mu^{2}}-1\right)z^{2}+\left(\frac{\omega^{2}}{\mu^{2}}-\frac{r_{b}}{r_{s}}\frac{k^{2}}{\mu^{2}}\right)\mu r_{s}z-\ell(\ell+1)\right]\psi(z)=0\, ,
\end{equation}
\end{widetext}
which is formally the Schrödinger equation for the hydrogen atom. Normal modes have $\frac{\omega^{2}}{\mu^{2}}-\frac{k^{2}}{\mu^{2}}-1<0$, which in Eq.~\eqref{eq:SchApprox} corresponds to having a negative binding energy. For such a mode to be normalizable the Coulomb potential in Eq.~\eqref{eq:SchApprox} must be attractive, $\frac{\omega^{2}}{\mu^{2}}-\frac{r_{b}}{r_{s}}\frac{k^{2}}{\mu^{2}}>0$. Combining these two conditions, we learn that for a KK bound state to exist one needs 
\begin{equation}
    p<\frac{R_{y}/\lambda}{\sqrt{\frac{r_{b}}{r_{s}}-1}}\equiv p_{\star}\, .
\end{equation}
Below, we will show that this is actually a necessary condition in the entire parameter space, and not only when $\mu r_{b}\ll1$. Modes with KK momentum $p\geq p_{\star}$ are necessarily radiated to infinity, so $p_{\star}$ can be understood as the KK ionization threshold. Just as in the hydrogen atom, imposing that solutions of Eq.~\eqref{eq:SchApprox} are regular as $z\to0$ and bound at infinity sets a quantization condition in $\omega$ which, at leading order in $\mu r_{s}$, gives
\begin{equation}
    \omega_{n_{\text{H}}}=\sqrt{\mu^{2}+k^{2}}\left[1-\frac{\left(\mu^{2}-(r_{b}/r_{s}-1)k^{2}\right)^{2}r_{s}^{2}}{4\left(\mu^{2}+k^{2}\right)}\frac{1}{2n_{H}^{2}}\right]\,,
\end{equation}
with $n_{\text{H}}=1,2,3...$ and $n_{\text{H}}>\ell$. This shows that normal modes of TSs have a regime that is precisely hydrogenic, thus allowing a natural interpretation of the system as a gravitational atom. This behavior is also present in the quasi-bound states of superradiant massive fields around rotating BHs~\cite{DetweilerGA,Baumann:2019eav}, although in that case the states are only metastable, while here they are truly stationary. Indeed, the only decaying mechanism these states have is through
electromagnetic and gravitational wave emission.
  
\section{The normal mode spectrum of Topological Stars}\label{Sec:NormalSpectrumTS}

In what follows, it will be useful to rewrite our variables in terms of dimensionless quantities, set by the TS's scale $r_{b}$,
\begin{equation}\label{eq:dimlessVars}
	\frac{r}{r_b}=\tilde{r},\quad\omega=\frac{\tilde{\omega}}{r_b},\quad\mu=\frac{\tilde{\mu}}{r_b},\quad \eta=\frac{r_b}{r_s},\quad k=\frac{\tilde{k}}{r_b}\,.
\end{equation}
In addition, we will find that the quantities
\begin{equation}\label{eq:paramsDef}
\begin{aligned}
    \epsilon&=\tilde{\omega}^2-\tilde{k}^2-\tilde{\mu}^2\,, \\ 
    \sigma&=\tilde{k}^2\,(1-\eta)+\tilde{\mu}^2\, , \\ 
    \varphi_{\tilde{k}}&=2\,\tilde{k}^2(1-\eta)+\sigma+\Lambda\,(1-\eta)\, ,
\end{aligned}
\end{equation}
are the ones governing the inner structure of the potential. Furthermore, from here onward, the focus is entirely on normal modes of the TS, which have a purely real frequency $\tilde{\omega}$ (see Section~\ref{ssec:normalProof}), implying that $\epsilon$, defined in Eq.~\eqref{eq:paramsDef}, is also purely real. With this, the radial equation~\eqref{eq:radeq} reads
\begin{align}\label{eq:potEq}
	\psi''(z)&+\tilde{r}^3\,(\tilde{r}-1)\,\eta^2\,Q(\tilde{r})\,\psi(z)=0,\notag\\
	Q(\tilde{r})&=\epsilon-V_{\rm eff},\notag\\ V_{\rm eff}&=\frac{\tilde{k}^2\,(\eta-1)}{\eta\,\tilde{r}\,(\tilde{r}-1)}+\frac{(\tilde{r}\,\eta-1)\,\Lambda}{\eta\,\tilde{r}^3}-\frac{\sigma}{\eta\,\tilde{r}},
\end{align}
where $d\tilde{z}/d\tilde{r}=1/(\tilde{r}-1)(\eta\,\tilde{r}-1)$.

The next sections are devoted to studying the space of normal mode solutions of Eq.~\eqref{eq:potEq}. These have a purely real frequency, $\tilde{\omega}$, as shown in Section~\ref{ssec:normalProof}. Besides the harmonic, KK and monopole numbers $\ell, p, N$, normal frequencies are also labeled by the overtone number $n$, although we will avoid writing those labels explicitly for simplicity.

\subsection{Existence and classification}\label{ss:Existence}

As stated below Eq.~\eqref{eq:paramsDef}, for the remainder of this work, $\epsilon$ will be considered purely real for normal modes of the TS. Thus, translating Eq.~\eqref{eq:QNMvsBS} to the notation here, we see that the boundary conditions for bounded modes require $\epsilon<0$, whereas for $\epsilon>0$, the solutions are QNMs, previously studied in~\cite{Heidmann:2023ojf,Bianchi:2023sfs}. Hence, with the scalar equation written in a Schrödinger-like way, as in Eq.~\eqref{eq:potEq}, we may immediately infer that normal mode solutions depend on the sign of $Q$. Regions where $Q>0$ correspond to ``classically allowed'' regions, where the field is allowed to oscillate. Regions with $Q<0$ correspond to ``classically forbidden'' regions, where bound states cannot be localized. Therefore, to have normal mode solutions, the potential must allow for at least one classically allowed region outside of the star, in which case, bound modes are characterized by $V_{\rm eff}<\epsilon<0$, as in the standard square potential problem in quantum mechanics.

With this in mind, we may proceed to verify that $\sigma>0$ is a necessary condition to have normal mode solutions. To this end, assume the opposite holds, namely $\sigma<0$. Since $\epsilon<0$  is required for the existence of normal modes, and $\Lambda\geq0$ by definition, together with $2>\eta>1$ and $\tilde{r}\geq1$, we see that $Q<0$ everywhere, so there can be no such modes. Hence, taking $\sigma>0$ with $\epsilon<0$, there is always at least one classically allowed region, where we expect normal mode solutions to exist. This existence condition is precisely Eq.~\eqref{eq:KKion}, which was derived in the hydrogenic limit, and emerged as a condition of ``attractiveness'' of the gravitational interaction.

We will investigate the structure of the spectrum both analytically and numerically. The case of a perturbation with no KK momentum, $\tilde{k}=0$, is significantly easier to characterize, and we will work out the mode classification in full detail by studying the potential $Q(\tilde{r})$ defined in Eq.~\eqref{eq:potEq}. The addition of KK momentum, $\tilde{k}$, carried by the scalar field will be covered afterwards, whereby most of the substantially more complicated technical work will be omitted for brevity.

\subsection{Zero-KK modes}\label{sec:ZeroKK}

\begin{figure}[t]
	\includegraphics[width=1\linewidth]{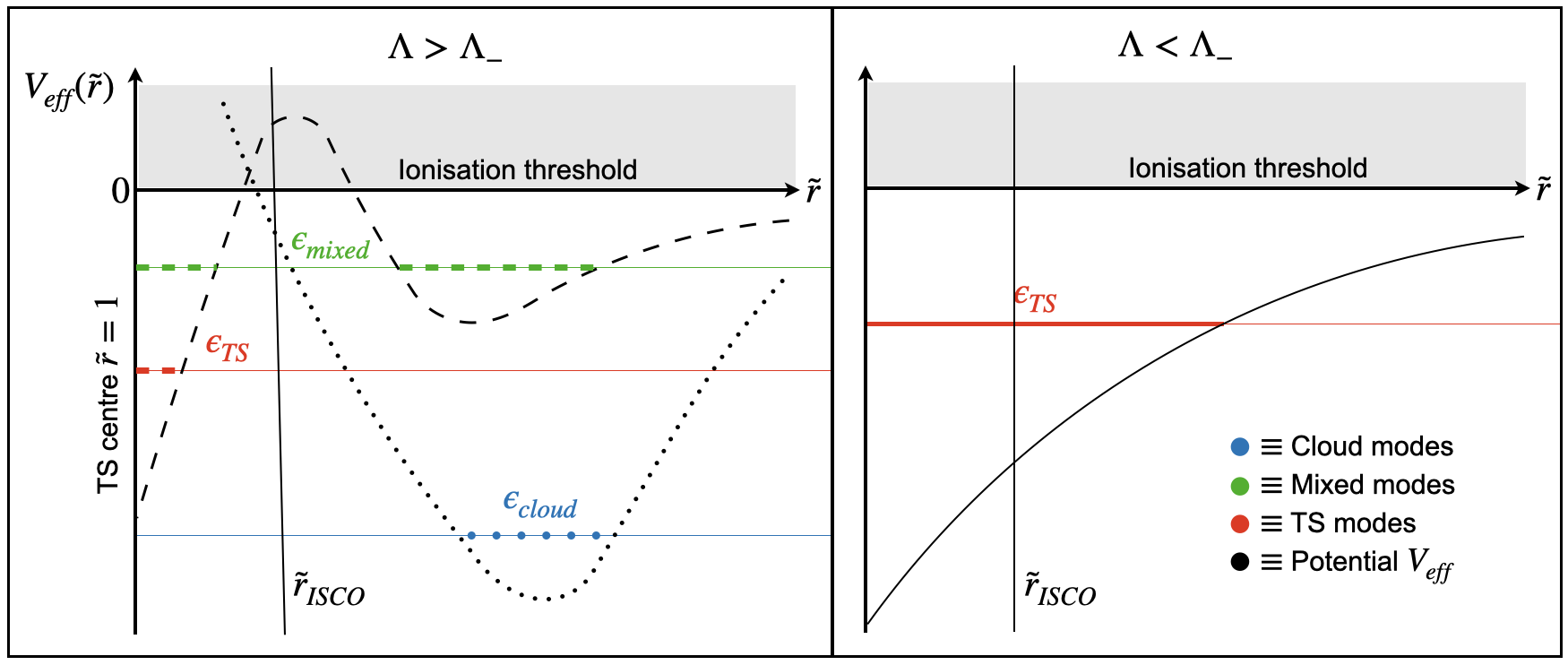} 
	\caption{Schematic representation of the scalar potential $V_{\rm eff}(\tilde{r})$ in Eq.~\eqref{eq:potEq}, and the classification of normal modes. Left panel: the case $\Lambda> \Lambda_{-}$. Given a normal mode energy level $\epsilon<0$, this can be a TS (red), mixed (green) or cloud (blue) mode. Dashed and dotted lines correspond to two illustrative configurations, where classically allowed regions are indicated with thick horizontal lines. The curve $V_{\rm eff}=0$ is the ionization threshold, since modes with $\epsilon>0$ are not bound. Right panel: Same, but for $\Lambda< \Lambda_{-}$, where only TS modes exist.}
	\label{Fig:Classification} 
\end{figure}

The scalar potential for zero-KK modes becomes
\begin{equation}\label{eq:Schrodinger_rad_eq_potential_k0_N}
	\hat{Q}\equiv Q_{\tilde{k}\to0}(\tilde{r})=\epsilon-\frac{(\tilde{r}\,\eta-1)\,\Lambda}{\eta\,\tilde{r}^3}+\frac{\tilde{\mu}^2}{\eta\,\tilde{r}}.
\end{equation}
Additionally, in the absence of KK momentum the existence condition $\sigma=\tilde{\mu}^2>0$ is always satisfied, and it is guaranteed that there will be at least one classically allowed region where normal modes can exist. It is precisely the structure of such regions that yields a sharp classification of bound states, as we discuss next. The extrema of the potential \eqref{eq:Schrodinger_rad_eq_potential_k0_N} are 
\begin{equation}\label{eq:KK0MaxMin_N}
	\tilde{r}_{\rm min}=\frac{3}{\eta\,(1+\alpha)},\quad\tilde{r}_{\rm max}=\frac{3}{\eta\,(1-\alpha)},
\end{equation}
where 
\begin{equation}
    \alpha\equiv\sqrt{1-\frac{\Lambda_-}{\Lambda}}\, , \quad \Lambda_-\equiv\frac{3\,\tilde{\mu}^2}{\eta^2}\, .
\end{equation}
One can distinguish two regimes, depending on whether the roots of $\hat{Q}'(\tilde{r})$, $\tilde{r}_{\rm min,max}$, are real or not: 
\begin{subequations}\label{eq:QDervKKZero}
	\begin{align}
		\Lambda < \Lambda_{-}:&\quad \text{$\hat{Q}$ has no local extrema,}\label{eq:noEx}\\
		\Lambda \geq \Lambda_{-}:&\quad \text{$\hat{Q}$ has one or two local extrema.}\label{eq:ex}
	\end{align}
\end{subequations}
The case $\Lambda < \Lambda_{-}$ corresponds to a low angular momentum regime: recall that $\Lambda=\ell(\ell+1)-(N/2)^{2}$. In this case, given a bound mode with energy $\epsilon$ (necessarily negative), there is a unique classically allowed region where the mode can be localized, which extends from the TS center $\tilde{r}=1$ to the turning point $r_{\rm turn}$ where $\hat{Q}(r_{\rm turn})=0$. The case $\Lambda \geq \Lambda_{-}$ is instead a moderate or large angular momentum regime, where a bound mode with energy $\epsilon$ can see either one, or two classically allowed regions disconnected by a centrifugal barrier. Such regions may contain the star's center, or may be disconnected from it and centered instead around $\tilde{r}_{\rm max}$. We notice that $\tilde{r}_{\rm max}$, whenever real, always lies outside the ISCO, $\tilde{r}_{\rm max}>\tilde{r}_{\rm ISCO}=3/\eta$, whereas $\tilde{r}_{\rm min}$, which always satisfies $0<\tilde{r}_{\rm min}\leq \tilde{r}_{\rm max}$, may even be outside the spacetime domain, $\tilde{r}_{\rm min}<1$. After these observations, we can establish the following classification based on the classically allowed regions for the modes, and their positions relative to the ISCO:
\begin{itemize}
    \item \textbf{TS modes}: only one classically allowed region whose maximum depth is within the ISCO.
     \item \textbf{Mixed modes}: two classically allowed regions.
    \item \textbf{Cloud modes}: only one classically allowed region, whose maximum depth is outside the ISCO.
\end{itemize}
This classification is illustrated schematically in Fig.~\ref{Fig:Classification}. 

Next, we describe in full detail where all types of modes exist in the parameter space. To this end, we introduce a quantity that is better suited for the upcoming analysis:
\begin{equation}\label{eq:nsEnCondk01}
	\gamma=\sqrt{-\epsilon}=\sqrt{\tilde{k}^2+\tilde{\mu}^2-\tilde{\omega}^2}.
\end{equation}
It can be viewed as the binding energy of the mode. We have $\epsilon<0$ for normal modes, hence, for them $\gamma\in\mathbb{R}$, otherwise, QNMs have imaginary $\gamma$. The limit $\gamma\to0$ corresponds to the ionization threshold between bound and unbound modes. Our strategy will be to first determine the regions where TS and cloud modes can possibly exist, and then look at their overlap to identify the mixed ones. It turns out that the phase space split given by the condition~\eqref{eq:QDervKKZero} provides a systematic way of carrying out this task. It is important to note that $\hat{Q}$ depends on the binding energy of the mode, hence the nature of a given classically allowed region might change depending on its energy.

Asymptotically, the potential approaches, for bound modes, a constant negative value:
\begin{equation}
	\lim\limits_{\tilde{r}\to\infty}\hat{Q}(\tilde{r})=\epsilon<0.
\end{equation}
Therefore, given that for $\tilde{k}=0$ we are always guaranteed at least one classically allowed region in the star's exterior, $\hat{Q}$ needs to have at least one positive-valued (local or global) maximum for $\tilde{r}\geq1$. We will investigate the presence of such regions in two steps, based on Eq.~\eqref{eq:QDervKKZero}. When no local extrema exist, as in Eq.~\eqref{eq:noEx}, one can show that the potential has a global maximum at the star surface, namely:
\begin{equation}\label{eq:starSurfacePot}
	\quad\hat{Q}(\tilde{r}=1)=\gamma_0^2-\gamma^2,
\end{equation}
where
\begin{equation}\label{eq:gamma0Def}
	\gamma_0=\sqrt{\frac{\varphi_0}{\eta}},\quad\varphi_0=\varphi_{\tilde{k}=0}.
\end{equation}
This is always located within the ISCO, $\tilde{r}_{\rm ISCO}=3/\eta$, and, as will become clear as we progress, it is the only possible classically allowed region where the TS family can be localized. 

The second case, corresponding to Eq.~\eqref{eq:ex}, requires us to also take into account the possible local extrema of $\hat{Q}$ given by:
\begin{equation}\label{eq:QminMax}
	\hat{Q}(\tilde{r}=\tilde{r}_{\rm min})=\gamma_-^2-\gamma^2,\quad\hat{Q}(\tilde{r}=\tilde{r}_{\rm max})=\gamma_+^2-\gamma^2,
\end{equation}
where
\begin{equation}\label{eq:gammasPmDefs}
	\gamma_{\pm}=\sqrt{\frac{(1\mp\alpha)\,(1\pm2\,\alpha)}{9\,(1\pm\alpha)}}\,\tilde{\mu},\quad\mbox{for}\quad\Lambda\geq\Lambda_-.
\end{equation}
In the range where the extrema exist, we have:
\begin{align}\label{eq:QminMaxRelative}
	\gamma_-&\leq\gamma_+,\quad\mbox{for}\quad\Lambda_-\leq\Lambda,\notag\\
	\hat{Q}_{\rm min}&\leq\hat{Q}_{\rm max},\quad\mbox{for}\quad\Lambda_-\leq\Lambda,
\end{align}
with $\hat{Q}(\tilde{r}=\tilde{r}_{\rm min/max})=\hat{Q}_{\rm min/max}$ and the equalities on the left are satisfied simultaneously with those on the right. Clearly, then $\hat{Q}_{\rm min}$ is a local minimum, whereas $\hat{Q}_{\rm max}$ is a local maximum. Their signs $\hat{Q}_{\rm min/max}$ change according to the binding energy of the mode as follows:
\begin{widetext}
		\begin{equation}\label{eq:QminMaxTable}
			\begin{tabular}{ |c||c|c|c| }
				\hline
				$\Lambda\geq\Lambda_-$ & $\gamma_+<\gamma$ & $0<\gamma_-<\gamma<\gamma_+$ & $0<\gamma<\gamma_-$ \\
				\hhline{|=#=|=|=|}
				$\hat{Q}_{\rm min}$ & <0 & <0 & >0 \\
				\hline
				$\hat{Q}_{\rm max}$ & <0 & >0 & >0 \\
				\hline
			\end{tabular}
		\end{equation}
\end{widetext}
While $\hat{Q}_{\rm min/max}$ do not exist for $\Lambda<\Lambda_-$, as stated below Eq.~\eqref{eq:QDervKKZero}, and, thus, cannot affect the structure of classically allowed regions in that regime, the other potential well, Eq.~\eqref{eq:starSurfacePot}, where states can possibly be localized, exists for:
\begin{align}\label{eq:lambdaHatDef}
	&\hat{Q}(\tilde{r}=1)>0,\quad\mbox{for}\quad\Lambda<\hat{\Lambda},\notag\\
	&\hat{\Lambda}=\frac{\tilde{\mu}^2}{\eta-1},\quad\mbox{and}\quad\Lambda_-<\hat{\Lambda},
\end{align}
with $\hat{\Lambda}$ defined through $\left.\gamma_0\right|_{\Lambda=\hat{\Lambda}}=0$, and the inequality on the second line holding for $1<\eta\leq2$ and all $\tilde{\mu}\neq0$. It is Eq.~\eqref{eq:lambdaHatDef} that allows us to deduce that there is a range of parameters where $\hat{Q}$ possesses two distinct, classically allowed regions with bound modes possibly having support over both of them.

As already stated, $\hat{Q}_{\rm max}$ is always located beyond the ISCO, $\tilde{r}_{\rm ISCO}=3/\eta$, however, the same is not true for $\hat{Q}_{\rm min}$. When the latter is negative, it acts like a centrifugal barrier between the two possible classically allowed regions, $\hat{Q}(\tilde{r}=1)$ and $\hat{Q}_{\rm max}$. However, when $\hat{Q}_{\rm min}>0$, represented by the last column of Table~\ref{eq:QminMaxTable}, we have to be more careful. Accordingly, we define
\begin{equation}
\Lambda_+=\frac{4\,\tilde{\mu}^2}{\eta^2},\quad\left.\gamma_-\right|_{\Lambda=\Lambda_+}=0.
\end{equation}
It is a simple exercise to check that $\Lambda_+<\hat{\Lambda}$ for all $\tilde{\mu}\neq0$ and $1<\eta\leq2$. We can thus conclude that in the regime
\begin{equation}
	0<\gamma<\gamma_-,\quad\Lambda_-\leq\Lambda<\Lambda_+,
\end{equation}
the mode energies are ``high'' enough that they have support over both extrema, $\hat{Q}_{\rm min/max}$, as well as over the classically allowed region near the star's surface. In fact, they are trapped in the whole region between the perfectly reflecting boundary at the star's surface and the mass barrier at infinity. In this case, deciding which family a mode belongs to reduces, with the help of Eq.~\eqref{eq:QminMaxRelative}, to comparing the depths of $\hat{Q}(\tilde{r}=1)$ and $\hat{Q}_{\rm max}$, or equivalently, $\gamma_0$ and $\gamma_+$, due to Eqs.~\eqref{eq:starSurfacePot} and \eqref{eq:QminMax}.

With that discussion in mind and in light of the sign distribution in Eq.~\eqref{eq:QminMaxTable}, it is at $\hat{Q}_{\rm max}$, when representing a classically allowed region, that cloud modes can be localized.

We can now look back at the definitions of the three mode families and together with Eqs.~\eqref{eq:starSurfacePot}, \eqref{eq:QminMax}, \eqref{eq:QminMaxTable} and \eqref{eq:lambdaHatDef} determine the regions where they exist. To that end, we also define:
\begin{gather}
	\Lambda_0=\frac{4\,\tilde{\mu}^2}{(\eta-1)\,(\eta+3)},\quad\left.\gamma_0\right|_{\Lambda=\Lambda_0}=\gamma_+,
\end{gather}
and point out the following relations:
\begin{align}
	\gamma_+<\gamma_0,\quad&\mbox{for}\quad\Lambda_-\leq\Lambda<\Lambda_0,\notag\\
	\gamma_0=\gamma_+,\quad&\mbox{for}\quad\Lambda=\Lambda_0,\notag\\
	\gamma_0<\gamma_+,\quad&\mbox{for}\quad\Lambda_0<\Lambda.
\end{align}
Then we proceed, one by one, beginning with:
\begin{gather}
	\mbox{TS modes:}\notag\\
	\begin{align}\label{eq:TSregion}
		&1<\eta<\frac{3}{2}\Rightarrow
		\begin{cases}
			0<\gamma\leq\gamma_0,\quad&\mbox{for}\quad\Lambda\leq\Lambda_-\\
			0<\gamma<\gamma_-,\quad&\mbox{for}\quad\Lambda_-\leq\Lambda<\Lambda_+\\
			\gamma_+<\gamma\leq\gamma_0,\quad&\mbox{for}\quad\Lambda_-\leq\Lambda\leq\Lambda_0
		\end{cases},\notag\\
		&\frac{3}{2}<\eta<2\Rightarrow
		\begin{cases}
			0<\gamma\leq\gamma_0,\quad&\mbox{for}\quad\Lambda\leq\Lambda_-\\
			0<\gamma<\gamma_-,\quad&\mbox{for}\quad\Lambda_-\leq\Lambda<\Lambda_0\\
			\gamma_+<\gamma\leq\gamma_0,\quad&\mbox{for}\quad\Lambda_-\leq\Lambda\leq\Lambda_0
		\end{cases},
	\end{align}
\end{gather}
followed by the transition region:
\begin{gather}
	\mbox{Mixed modes:}\notag\\
	\begin{align}\label{eq:mixedRegion}
		&1<\eta<\frac{3}{2}\Rightarrow
		\begin{cases}
			\gamma_-\leq\gamma\leq\gamma_+ & \Lambda_-<\Lambda\leq\Lambda_+\\
			0<\gamma\leq\gamma_+ & \Lambda_+\leq\Lambda\leq\Lambda_0\\
			0<\gamma\leq\gamma_0 & \Lambda_0\leq\Lambda<\hat{\Lambda}
		\end{cases},\notag\\
		&\frac{3}{2}<\eta<2\Rightarrow
		\begin{cases}
			\gamma_-\leq\gamma\leq\gamma_+ & \Lambda_-<\Lambda\leq\Lambda_0\\
			\gamma_-\leq\gamma\leq\gamma_0 & \Lambda_0\leq\Lambda\leq\Lambda_+\\
			0<\gamma\leq\gamma_0 & \Lambda_+\leq\Lambda<\hat{\Lambda}
		\end{cases},
	\end{align}
\end{gather}
and finally:
\begin{gather}
	\mbox{Cloud modes:}\notag\\
	\begin{align}\label{eq:cloudRegion}
		&1<\eta<\frac{3}{2}\Rightarrow
		\begin{cases}
			\gamma_0<\gamma\leq\gamma_+,\quad&\mbox{for}\quad\Lambda_0<\Lambda\leq\hat{\Lambda},\\
			0<\gamma\leq\gamma_+,\quad&\mbox{for}\quad\hat{\Lambda}\leq\Lambda
		\end{cases},\notag\\
		&\frac{3}{2}<\eta<2\Rightarrow
		\begin{cases}
			\gamma_0<\gamma\leq\gamma_+,\quad&\mbox{for}\quad\Lambda_0<\Lambda\leq\hat{\Lambda},\\
			0<\gamma\leq\gamma_-,\quad&\mbox{for}\quad\Lambda_0<\Lambda\leq\Lambda_+,\\
			0<\gamma\leq\gamma_+,\quad&\mbox{for}\quad\hat{\Lambda}\leq\Lambda
		\end{cases}.
	\end{align}
\end{gather}
The splits in two parts based on the value of $\eta$ come about because $\Lambda_+=\Lambda_0$ for $\eta=\frac{3}{2}$, the transition point between TSs of the first and second kind. Hence, for $1<\eta<\frac{3}{2}$, $\Lambda_0$ is to the right of both $\Lambda_{\pm}$, whereas, for $\frac{3}{2}<\eta\leq2$ it sits between them, as $\Lambda_0>\Lambda_-$ as long as $\tilde{\mu}\neq0$.

We proceed with some comments on the different regions, starting with the TS modes. The region which is always localized near the star surface, $\tilde{r}=1$, is finite for any non-infinite and non-zero scalar mass, $\tilde{\mu}$, as well as non-extremal value of $2\geq\eta>1$. This can be seen from the requirement that $\varphi>0$ for $\gamma_0$ to be real, which puts a bound on $\ell$ in terms of the field's mass. There are two exceptions to this: the extremal limit, $\eta=1$ and $\Lambda=0$ -- that is, $\ell=0$. The latter implies that for any non-zero scalar mass, the $\ell=0$ mode's radial function will have at least some support in this region. The former will be investigated separately because the inner boundary condition changes. In Appendix~\ref{secA:quasimodes2} we give an approximate description of the TS modes using quasimodes~\cite{holzegel2014quasimodes,Keir:2014oka} in the regime $\tilde{\mu}\gg1$. The resulting frequency spectrum is given by Eq.~\eqref{eq:largeMuApprox}.

The cloud modes' region, on the other hand, closes off only asymptotically, because
\begin{equation}\label{eq:gammaPAsymp}
	\lim\limits_{\ell\to\infty}\gamma_+=0.
\end{equation}
It is centered on $\tilde{r}_{\rm max}$, which, in the eikonal limit, scales as:
\begin{equation}\label{eq:cloudsPosEikonal}
	\lim\limits_{\ell\to\infty}\tilde{r}_{\rm max}=\frac{2\,\eta\,\Lambda}{\tilde{\mu}^2}=\frac{2\,\eta\,\ell(\ell+1)}{\tilde{\mu}^2}.
\end{equation} 
As for the TS modes, we have provided an approximate construction of the cloud modes using quasimodes in Appendix~\ref{secA:quasimodes}, but this time valid for $\ell\gg1$. It recovers Eq.~\eqref{eq:cloudsPosEikonal}, which also matches the geodesic analysis in Section~\ref{ssec:geoAnalysis}. The frequency spectrum is given by Eq.~\eqref{eq:eikonalApprox}. It has qualitatively the same form as the real part of the frequency\footnote{In fact, we have done the computation for $0<\eta<1$, determining both the real and imaginary parts of the frequency in the eikonal limit, and one can reduce the result all the way to 4D Schwarzschild.} in the black string case ($0<\eta<1$) for $\ell\gg1$. In fact, the cloud modes' properties in the eikonal regime exactly parallel those of neutral scalar clouds around static, spherically symmetric, asymptotically flat BHs in 4D.

Finally, we note that for $\eta=2$, $\hat{\Lambda}=\Lambda_+$, however, $\Lambda_0>\Lambda_-$ for $2\geq\eta>1$ (they are equal at $\eta=3$), indicating that the mixed region never vanishes in our range of parameters. Nevertheless, this does not necessarily imply that one will always find such a mode in the spectrum for a given $\ell,\,p,\,N,\,n$, and, indeed, we have found examples of such situations. On the other hand, when a mixed mode exists, the rough distribution of its radial function between the two classically allowed regions depends on the relative height of $\hat{Q}(\tilde{r}=1)=\gamma_0^2-\gamma^2$ and $\hat{Q}_{\rm max}=\gamma_+^2-\gamma^2$, that is, on $\gamma_0^2-\gamma_+^2$, which vanishes for $\alpha=\frac{3-\eta}{2\,\eta}$.

\begin{figure}[t]
	\centering
	\includegraphics[width=\linewidth]{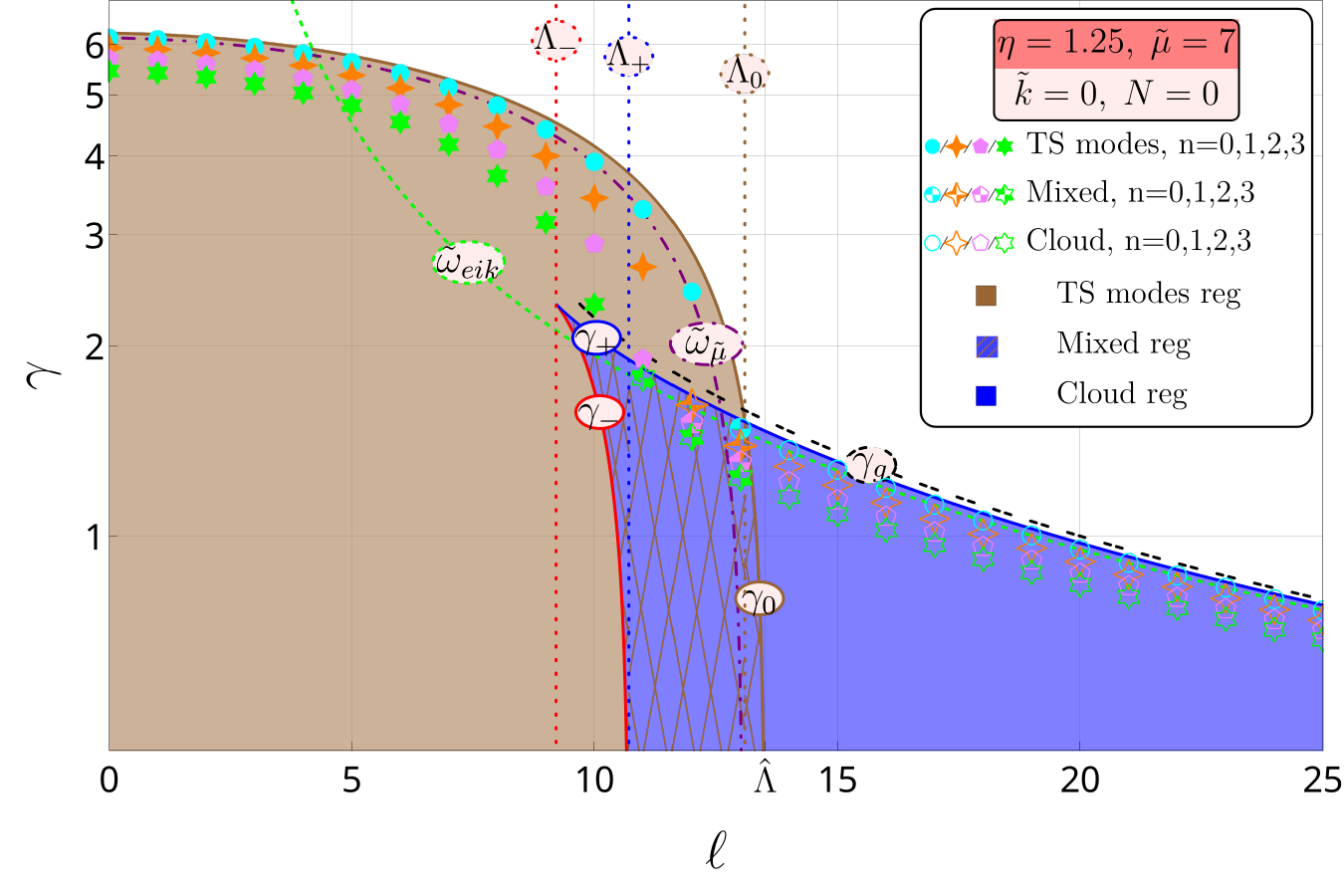}
	\vspace*{-6mm}
	\caption{Binding energy $\gamma$ of massive scalar field modes as a function of $\ell$ for a TS, as defined in Eq.~\eqref{eq:nsEnCondk01}, and plotted on a log scale for clarity. We set $\eta=1.25$, $\tilde{\mu}=7$, $\tilde{k}=0$, $N=0$, and evaluate $\gamma$ numerically for the $n=0,\,1,\,2,\,3$ radial overtones (cyan disks, orange diamonds, pink pentagons and green six-pointed stars, respectively). Filled markers are pure TS modes. Empty markers correspond to cloud modes, localized outside of the timelike geodesics ISCO. Half-filled markers indicate mixed modes. The regions where different modes can exist are shaded differently for clarity. The dashed green line for the eikonal approximation, Eq.~\eqref{eq:eikonalApprox}, and the dashed-dotted magenta line for the large scalar mass approximation, Eq.~\eqref{eq:largeMuApprox}, are for $n=0$ only. In black we plot the geodesic approximation $\gamma_g=\sqrt{\tilde{\mu}^2\,\big(1-\mathcal{E}^2\big)}$, with $\mathcal{E}$ from Eq.~\eqref{eq:circOrbits}.}
	\label{fig:k0mu7eta125Regs}
\end{figure}
\begin{figure}[!h]
	\centering
	\includegraphics[width=\linewidth]{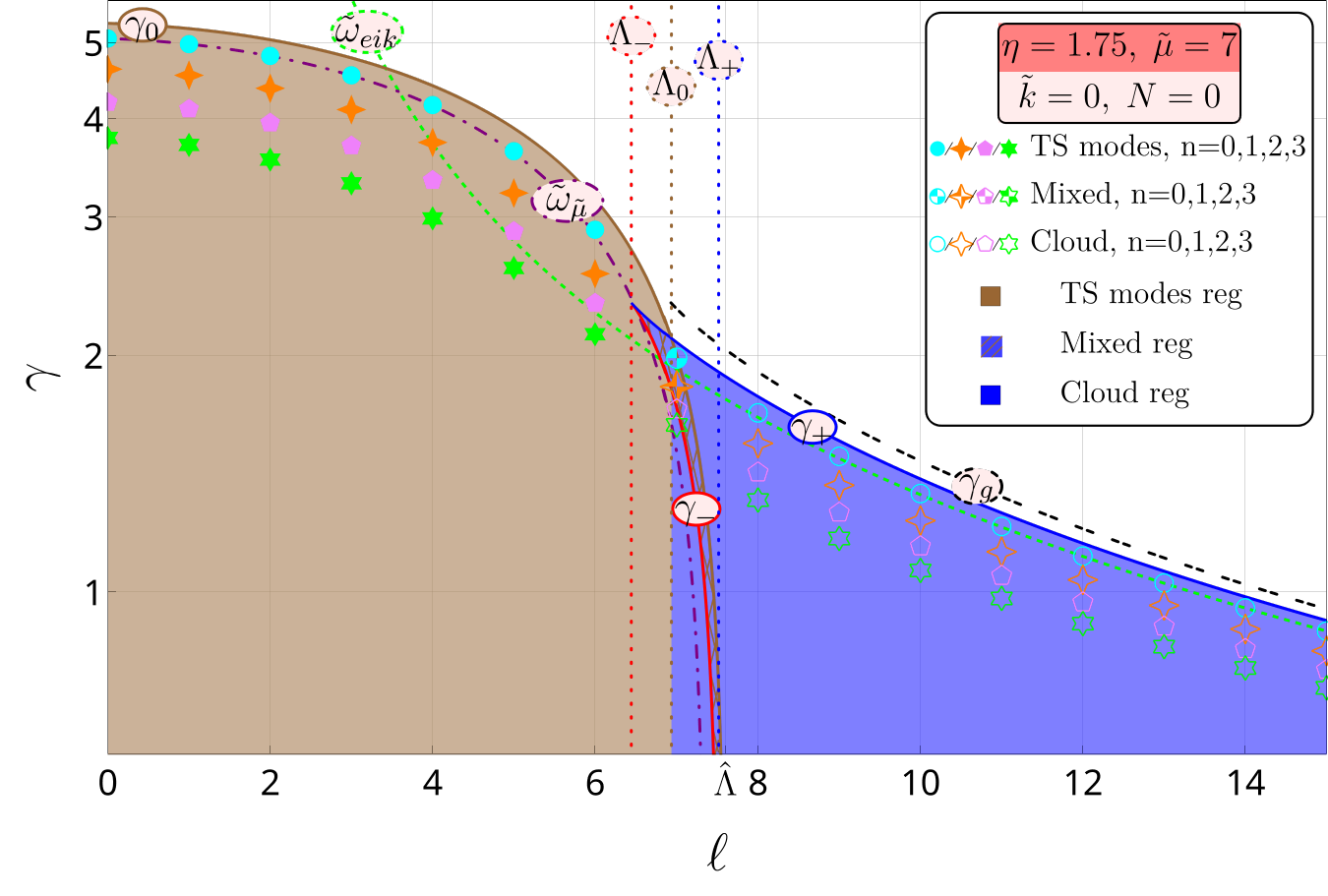}
	\vspace*{-6mm}
	\caption{Same as Fig.~\ref{fig:k0mu7eta125Regs}, but for $\eta=1.75$, $\tilde{\mu}=7$, $\tilde{k}=0$, $N=0$. Note the change in order between $\Lambda_0$ and $\Lambda_+$ with respect to Fig.~\ref{fig:k0mu7eta125Regs}, as anticipated by the analysis in the main text (as well as the different range on the horizontal axis).}
	\label{fig:k0mu7eta175Regs}
\end{figure}

In Figs.~\ref{fig:k0mu7eta125Regs} and \ref{fig:k0mu7eta175Regs} we showcase the validity of our analysis for two different representative sets of parameters in the case of a TS of the second and first kind, respectively. In both cases we see that the $\ell=0$, $n=0$ mode is the one with the highest binding energy $\gamma$ (i.e., the lowest frequency $\tilde{\omega}$). As we increase $\ell$, but keep the radial overtone number fixed, the energy decreases (the frequency increases) and from the eikonal approximation of Eq.~\eqref{eq:eikonalApprox}, we know that it approaches zero asymptotically (the frequency tends to the scalar mass $\tilde{\mu}$ in the same limit, for $\tilde{k}=0$). Moreover, the $n=0$ modes almost saturate the bounds on their binding energies in all three regions ($\gamma_0$ for TS modes, and $\gamma_+$ for mixed and cloud modes). As the overtone number $n$ is increased, while keeping $\ell$ fixed, the energies decrease, faster for the TS modes in comparison to the cloud ones (for mixed modes the rate depends on which type dominates the ``mixture''). We expect that for large overtones, $n\gg1$, the binding energy will tend to zero, $\gamma\to0$. This picture holds qualitatively for all scalar masses $\tilde{\mu}$ and for all $\eta$'s that we have probed, except for $\sigma\to0$ (for $\tilde{k}=0$ this is equivalent to $\tilde{\mu}\ll1$), when the TS region shrinks to an infinitesimal size.

Finally, in the plots, we have also included the binding energy associated to the circular timelike geodesics from Section~\ref{ssec:geoAnalysis}: $\gamma_g=\sqrt{\tilde{\mu}^2\,\big(1-\mathcal{E}^2\big)}$, with $\mathcal{E}$ given in Eq.~\eqref{eq:circOrbits}. The analysis in Section~\ref{ssec:geoAnalysis} holds to leading order in the eikonal limit, and, as expected, $\gamma_g$ agrees with $\gamma_+$ to that order for $\ell\gg1$.

\begin{figure*}[t]
	\label{fig:eta125k0wavefunctions}
	\centering
	\includegraphics[width=\linewidth]{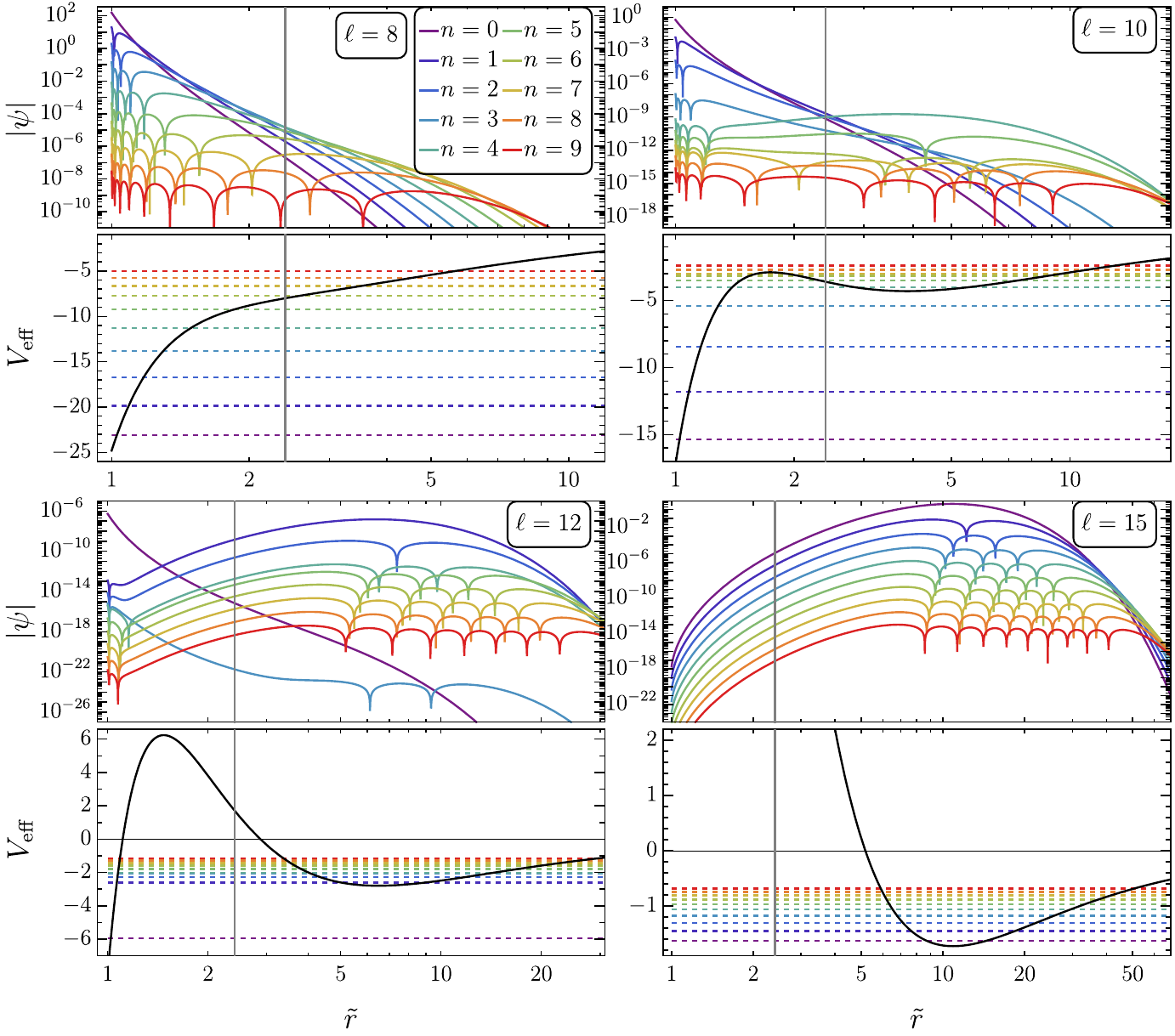}
	\caption{Top Panels: Scalar wavefunctions on the TS background with parameters as in Fig.~\ref{fig:k0mu7eta125Regs}, with increasing overtone number. Bottom Panels: $V_{\rm eff}$ (Black line) for the chosen parameters compared with overtone ``energies'' $\epsilon$ (colored lines). We take $\ell=8$, $\ell=10$, $\ell=12$, $\ell=15$, corresponding to the $\ell<\Lambda_-$, $\Lambda_-<\ell<\Lambda_+$, $\Lambda_+<\ell<\Lambda_0$, and $\Lambda_0<\ell$ regions of parameter space, respectively. The gray line corresponds to the ISCO radius $3r_S$. Modes with low $\ell$ are always localized close to the star. Waves are localized further from the star's surface as both overtone number and $\ell$ values increase, in line with the mode classification scheme in Eqs. \eqref{eq:TSregion}--\eqref{eq:cloudRegion}.}
\end{figure*}

\begin{figure*}[t]
	\label{fig:eta175k0wavefunctions}
	\centering
	\includegraphics[width=\linewidth]{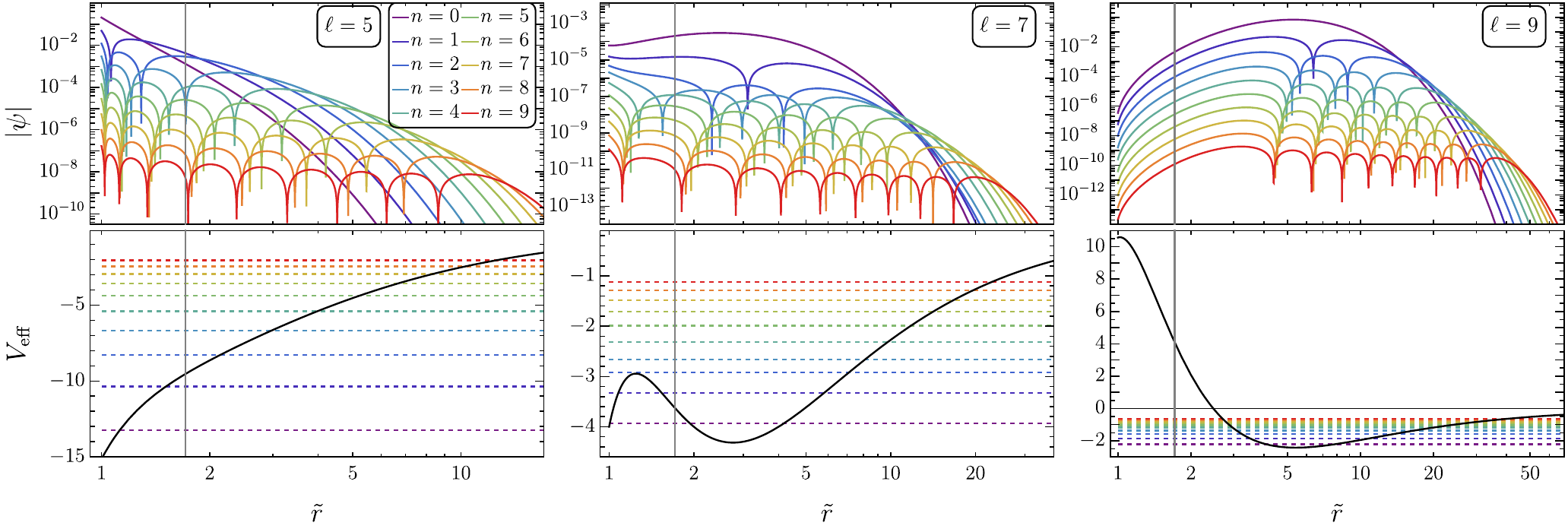}
	\caption{Same as Fig.~\ref{fig:eta125k0wavefunctions}, but for the parameters in Fig.~\ref{fig:k0mu7eta175Regs}. From left to right, we take $\ell=5$, $\ell=7$, $\ell=9$, corresponding to the $\ell<\Lambda_-$, $\Lambda_0<\ell<\Lambda_+$, and $\Lambda_+<\ell$ regions, respectively. Note that there is no $\ell$ value which lies within the $\Lambda_-<\ell<\Lambda_0$ region for the parameters chosen in Fig.~\ref{fig:k0mu7eta175Regs}.}
\end{figure*}

To further illustrate the behavior of the field as mode and star parameters are varied, in Figs.~\ref{fig:eta125k0wavefunctions} and \ref{fig:eta175k0wavefunctions} we plot the wavefunction $\psi(\tilde{r})$ and their corresponding energy levels $\epsilon$, superimposed on the effective potential $V_{\rm eff}$, for various regimes of the parameter space. Solutions of the radial equation \eqref{eq:radeq} were found by employing the standard continued fraction method of Leaver~\cite{Leaver:1985ax}, imposing the appropriate boundary conditions, i.e. Eqs. \eqref{eq:rb} and \eqref{eq:rinf}. In addition, recall from Section~\ref{ss:Existence} that the presence of a classically allowed region for bounded modes, $Q>0$, is equivalent to $V_{\rm eff}<\epsilon<0$. In Fig.~\ref{fig:eta125k0wavefunctions} we use the same parameters as Fig.~\ref{fig:k0mu7eta125Regs} for increasing $\ell$ values. Taking $\ell=8$, where we expect to only have TS modes, we see that the potential always only admits a single classically allowed region. As the overtone index increases, the scalar wave is simply localized further away from the star, but always retains the same behavior. The situation is different for the $\ell=10$ case. For the first few overtones, there is only one classically allowed region close to the star, so the waves behave as TS modes. As the overtone ``energy'' $\epsilon=-\gamma^2$ moves past the $\gamma_+$ threshold, see \eqref{eq:QminMaxTable}, a second classically allowed region develops, and the modes become mixed, being localized both close to the star and near $\hat{Q}_{\rm max}$. Eventually, the overtone reaches the $\gamma_-$ threshold, so that the wave sees a single classically allowed region again. The waves here go back to behaving like TS modes, and just spread out over the whole range of the classically allowed region. Distinctly, for $\ell=12$, modes start out as TS modes with only one classically allowed region, and eventually develop a second one where the modes are localized. As overtone increases, the modes stay localized far away from the star near the second minimum of the potential (we talk about minima, instead of maxima, because we are plotting $-\hat{Q}$). Lastly, in the $\ell=15$ region where we expect only cloud modes to exist, as again a single classically allowed region is present -- which is the potential's minimum outside of the ISCO -- the waves are localized there and spread out further as the overtone number is increased. 

For the $\eta=1.75$ case in Fig.~\ref{fig:eta175k0wavefunctions}, the picture is essentially the same, albeit with some slight qualitative differences in the $\ell=7$ mixed-mode case, due to the second potential minimum being deeper than the potential's value at the star's surface.


\begin{figure*}[t]
	\label{fig:k0_wavefunctions_increasing_ell}
	\centering
	\includegraphics[width=\linewidth]{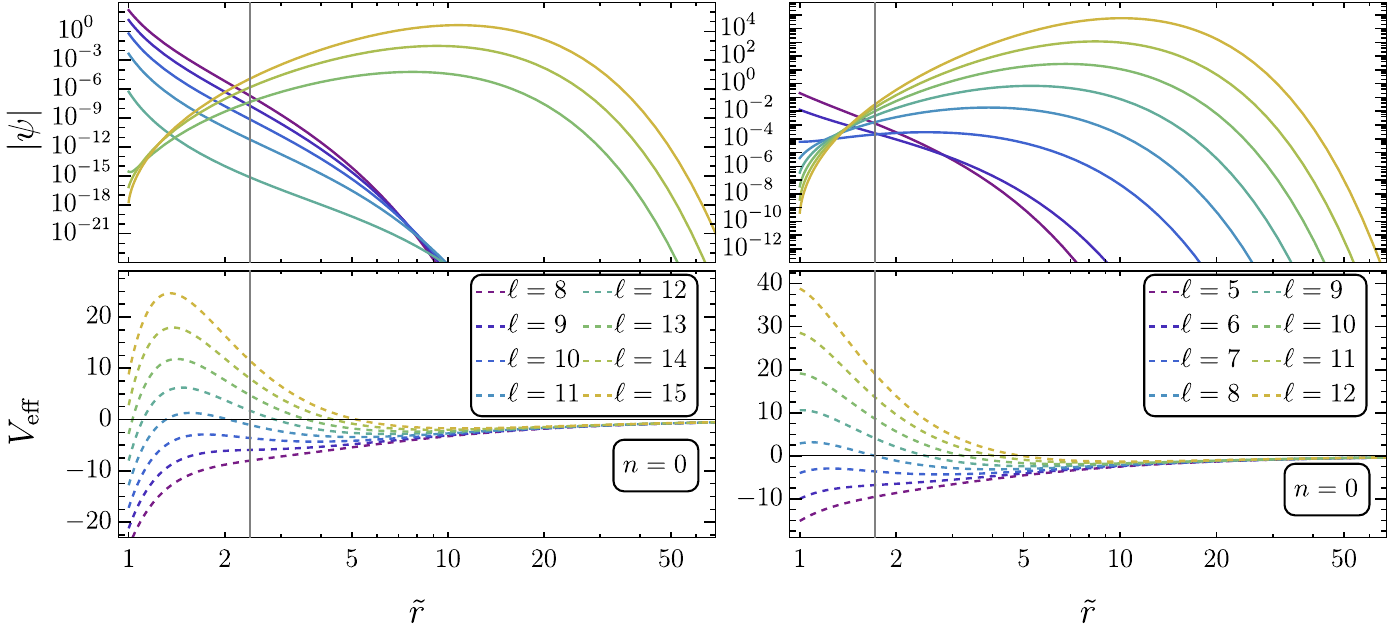}
	\caption{Left: Wavefunction and potential for the fundamental mode, using the same parameters as in Fig.~\ref{fig:k0mu7eta125Regs} and varying $\ell$. Near the transition region at $\ell\sim9$, the wavefunctions begin to move outwards towards the cloud, and eventually localize in the cloud once the $\ell>\Lambda_+$ threshold is reached. Right: Same as the left panel, for for the parameters used in Fig.~\ref{fig:k0mu7eta175Regs}. In both figures, the location of the ISCO radius is represented by a gray vertical line.}
\end{figure*}

The transitions from TS modes to mixed modes and cloud modes as the angular momentum increases is further demonstrated in Fig.~\ref{fig:k0_wavefunctions_increasing_ell}. We show how the fundamental mode wavefunctions change as $\ell$ increases near the threshold region $\ell\sim\Lambda_-$ for the same parameters as in Figs.~\ref{fig:k0mu7eta125Regs} and \ref{fig:k0mu7eta175Regs}. In both cases, we see the waves localize near the star, and gradually transition to localizing in the cloud region. Physically, this may be explained by noting that larger angular momentum in the field serves to push the field further outwards, towards the region where the field exhibits massive geodesic behavior.

\begin{figure}[t]
	\centering
	\includegraphics[width=\linewidth]{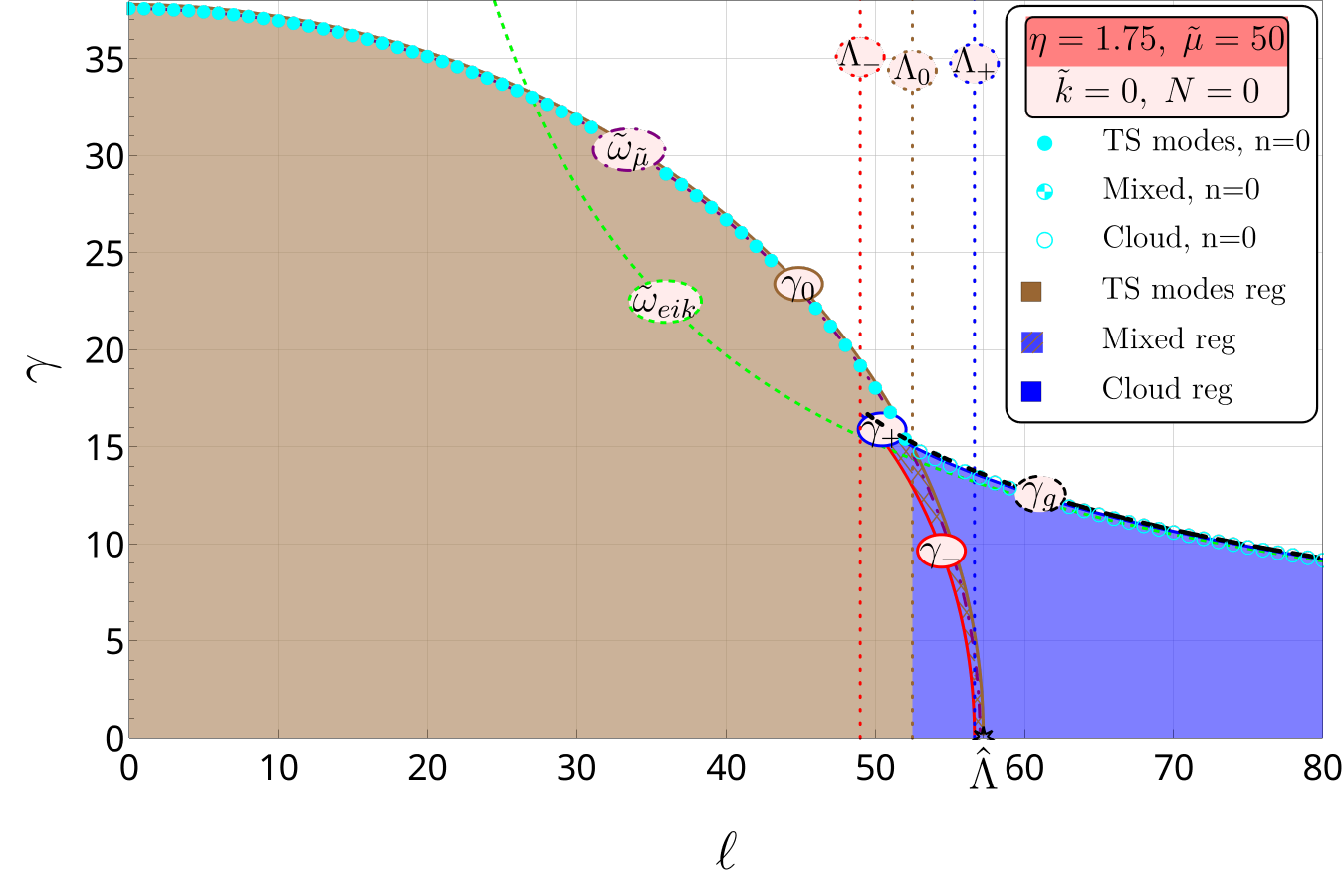}
	\vspace*{-6mm}
	\caption{Fundamental normal mode spectrum in the regime $\mu r_{b}\gg1$. The circular geodesic prediction~\eqref{eq:CircularPred}, represented in black by $\gamma_g=\sqrt{\tilde{\mu}^2\,\big(1-\mathcal{E}^2\big)}$ with Eq.~$\mathcal{E}$ from \eqref{eq:circOrbits}, describes well the behavior of cloud modes, and terminates with good precision where cloud modes transition into TS modes.}
	\label{fig:k0mu50eta175Regs}
\end{figure}

As anticipated in the analysis of Section~\ref{ssec:geoAnalysis}, in the limit $\mu r_{b}\gg1$ the spectrum is determined by timelike geodesics. There, it was shown that for large enough $\ell$, in particular $\ell\gtrsim \sqrt{3} \mu r_{s}$, the mode frequencies are dictated by the energy of stable circular geodesics
\begin{equation}\label{eq:CircularPred}
    \omega=\mu \ \mathcal{E}\left(\frac{\ell}{\mu r_{b}} \right)\, ,
\end{equation}
where the function $\mathcal{E}(x)$ is given in Eq.~\eqref{eq:circOrbits}. This family terminates at $\ell\sim \sqrt{3} \mu r_{s}$, below which the solutions are no longer localized around circular geodesics, and modes probe the interior of the ISCO. This behavior is illustrated in Fig.~\ref{fig:k0mu50eta175Regs}, where the geodesic prediction \eqref{eq:CircularPred} is contrasted with accurate numerical results for $\mu r_{b}=50$, which we expect to be well within the regime $\mu r_{b}\gg1$. One can see that the curve given by \eqref{eq:CircularPred} follows very well the numerical result and terminates at $\ell= \sqrt{3} \mu r_{s}$, which is with good precision the angular momentum where the solutions change behavior from cloud to TS modes. 

\begin{figure}[t]
	\label{fig:hydrogenic_spectrum}
	\centering
	\includegraphics[width=\linewidth]{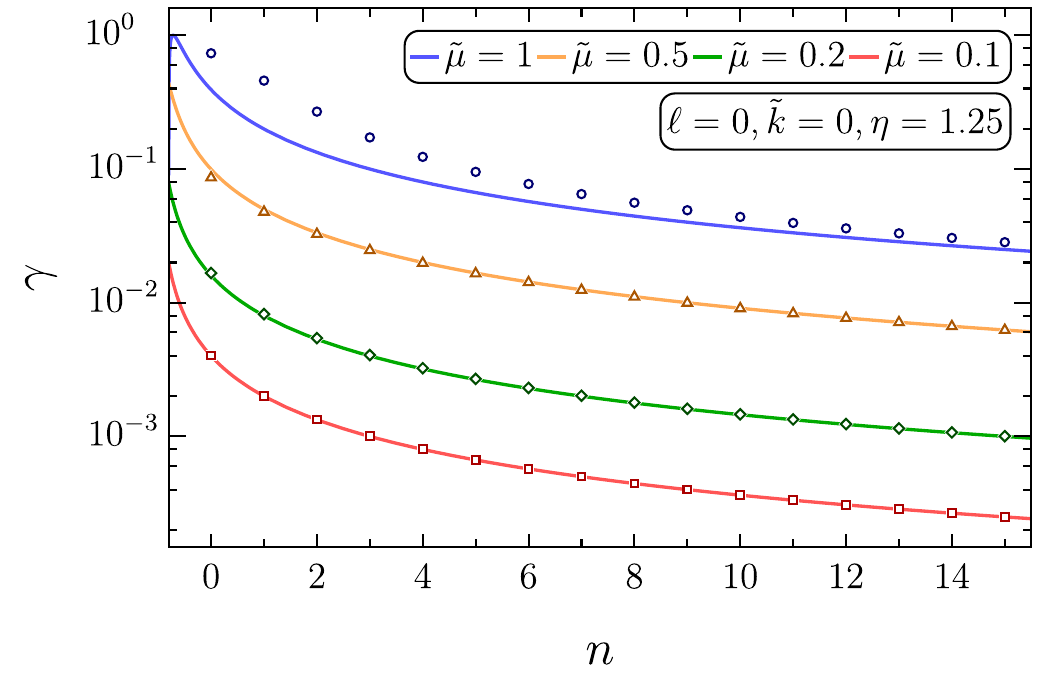}
	\caption{Comparison of normal mode spectra for various choices of $\tilde{\mu}$, in the hydrogenic limit $\tilde{\mu}\ll1$. Solid lines are the analytic prediction of Eq.~\eqref{eq:hydrogenic_spectrum}, and points are the numerical results generated through Leaver's method. We find good agreement with the predictions of Eq.~\eqref{eq:hydrogenic_spectrum}, which improves as the value of $\tilde{\mu}$ is made smaller. Here we set $\eta=1.25$, $\ell=\tilde{k}=0$ to demonstrate the behavior of the spectrum, but we find qualitatively similar results for nonzero choices of KK momentum and for other values of $\eta$ and $\ell$.}
\end{figure}

Finally, in Fig.~\ref{fig:hydrogenic_spectrum} we compare the hydrogenic ($\tilde{\mu}\ll1$) limit prediction in Eq.~\eqref{eq:hydrogenic_spectrum} to numerical results obtained through Leaver's method, setting $k=\ell=0$. We find good agreement with the prediction as $\tilde{\mu}$ decreases. We find qualitatively similar results for other choices of $k,\ell,$ and $r_b$. 

\subsection{KK modes}\label{sec:KK}

Allowing the scalar field to carry KK momentum does not lead to any major qualitative changes from the picture outlined in the previous subsection. However, looking at the definition of $Q(\tilde{r})$ in \eqref{eq:potEq}, we see that performing a similar analysis is obstructed by the increase in degree in $\tilde{r}$ of the polynomial we have to work with. It turns out that it is slightly easier to include the prefactors and consider:
\begin{equation}\label{eq:kPot}
	Q_k(\tilde{r})=\tilde{r}^3\,(\tilde{r}-1)\,\eta^2\,Q(\tilde{r}).
\end{equation}
One can then technically replicate what was done for $\tilde{k}=0$, however, we also need to look at the second and third derivatives of Eq.~\eqref{eq:kPot} and almost all steps of the process are more complicated with more subcases to consider, resulting in a lot of cumbersome expressions. We will therefore only mention two conditions that nonetheless allow one to, at least, distinguish the pure cloud modes from the rest. 

The potential~\eqref{eq:kPot} approaches a negative value at both ends of the $\tilde{r}$ domain. Combining this with the fact that we can have a maximum of four zero-crossings with the horizontal axis, we can deduce that there can be either one or two classically allowed regions present for any set of parameters. This can be accompanied by a study of the sign of the first derivative at $\tilde{r}=1$ and the second derivative of Eq.~\eqref{eq:kPot}, which is a quadratic in $\tilde{r}$. The former being negative allows for cloud modes only, as due to the asymptotics of the potential~\eqref{eq:kPot} we need at least 2 extrema to have a single classically allowed region. More are not possible because with the maximum number of extrema for $Q_k$, which is 3, a second classically allowed region and $Q'_k(\tilde{r}=1)<0$ require at least 2 minima and 2 maxima in the exterior of the star. Furthermore, determining the zeros of the second derivative will tell us how many concave regions (up or down) $Q_k(\tilde{r})$ has. The second classically allowed region only exists if there is more than one concavity, which implies that the discriminant of $Q_k''(\tilde{r})$ has to be non-negative. In the end, we derive the following conditions:
\begin{align}
	\mbox{TS modes possible:}&\;\;0<\gamma<\gamma_{\tilde{k}},\notag\\
	\mbox{Cloud modes possible:}&
	\begin{cases}
		0<\gamma<\frac{\sigma_\eta}{3},\;\;\Lambda_{\tilde{k}}<\sigma_\eta,\\
		0<\gamma<\gamma_c,\;\;\,\,\sigma_\eta<\Lambda_{\tilde{k}}.
	\end{cases}
\end{align}
where,
\begin{align}
	\gamma_c&=\sqrt{\frac{1}{3}\,\Big(4\,\Lambda_{\tilde{k}}+\sigma_\eta-2\,\sqrt{2\,\big(2\,\Lambda_{\tilde{k}}-\sigma_\eta\big)\big(\Lambda_{\tilde{k}}+\sigma_\eta\big)}\Big)},\notag\\
	\gamma_{\tilde{k}}&=\sqrt{\frac{\varphi_{\tilde{k}}}{\eta}},\quad\Lambda_{\tilde{k}}=\tilde{k}^2\,\frac{(\eta-1)}{\eta}+\Lambda,\quad\sigma_\eta=\frac{\sigma}{\eta},
\end{align}
$\varphi_{\tilde{k}}$ is given by Eq.~\eqref{eq:paramsDef}, implying $\gamma_{\tilde{k}=0}=\gamma_0$, $\Lambda_{\tilde{k}}\geq0$ for any choice of parameters, and $\Lambda_{\tilde{k}=0}=\Lambda$. The reality of the square root inside $\gamma_c$ requires $\Lambda_{\tilde{k}}>\sigma_\eta/2$, which is always satisfied in the regime of interest\footnote{Note that $\gamma_c$ does not reduce to $\gamma_+$ for $\tilde{k}=0$, because we derive the two bounds in different ways. In particular, we expect that the equivalent of $\gamma_+$ for non-zero $\tilde{k}$ will provide a more stringent bound than $\gamma_c$.}. The two regions thus defined can overlap and the area where that happens can host any type of mode -- pure or mixed. Further analysis of $Q_k$ and its derivatives can allow one to precisely characterize how that occurs, but the conditions are hard to work with. On the other hand, we can easily define the regions where only one set of modes exists:
\begin{align}
	\mbox{Only TS modes:}&\;\;0<\gamma<\gamma_{\tilde{k}},\;\;\Lambda_{\tilde{k}}<\sigma_\eta,\notag\\
	\mbox{Only cloud modes:}&
	\begin{cases}
		\gamma_{\tilde{k}}<\gamma<\frac{\sigma_\eta}{3},\;\;\Lambda_{\tilde{k}}<\sigma_\eta,\\
		\gamma_{\tilde{k}}<\gamma<\gamma_c,\;\;\,\,\sigma_\eta<\Lambda_{\tilde{k}}.
	\end{cases}
\end{align}
It is possible for $\gamma_{\tilde{k}}>\sigma_\eta/3$ when $\Lambda_{\tilde{k}}<\sigma_\eta$. When that happens, that part of the cloud region is not present.

There are two differences with the $\tilde{k}=0$ scenario. First,
\begin{equation}
	Q_k(\tilde{r}=1)=-\tilde{k}^2(\eta-1)\,\eta,
\end{equation} 
which is always negative, implying that the classically allowed region nearer to the star is always some non-zero, possibly infinitesimal, distance away from its surface. In this way, the KK momentum acts like a potential barrier, pushing the modes away from the star, with higher KK momentum values resulting in the modes being pushed further away. Note that at infinity, it still acts as an effective mass.

Second, we have
\begin{equation}\label{eq:noTSCond}
	\tilde{k}>\sqrt{\frac{\sigma}{2\,(\eta-1)}}\quad\Rightarrow\quad\varphi_{\tilde{k}}<0,
\end{equation}
leading to $\gamma_{\tilde{k}}$ being complex and, thus, the TS region disappearing for any $\ell$. This is in contrast to the $\tilde{k}=0$ case, where at least for $\ell=0$ we are always guaranteed some non-zero area for the TS region, as discussed below Eq.~\eqref{eq:gammaPAsymp}. Taking the definition of $\sigma$, Eq.~\eqref{eq:paramsDef}, and the requirement that it is positive for the existence of bound modes, allows us to rewrite Eq.~\eqref{eq:noTSCond} in terms of the scalar mass:
\begin{equation}\label{eq:noTSCond2}
	\frac{\tilde{\mu}}{\sqrt{\eta-1}}>\tilde{k}>\frac{\tilde{\mu}}{\sqrt{3\,(\eta-1)}}\quad\Rightarrow\quad\varphi_{\tilde{k}}<0.
\end{equation}

\begin{figure*}[t]
	\label{fig:eta175k191215}
	\centering
	\subfigure{\includegraphics[width=0.501\textwidth]{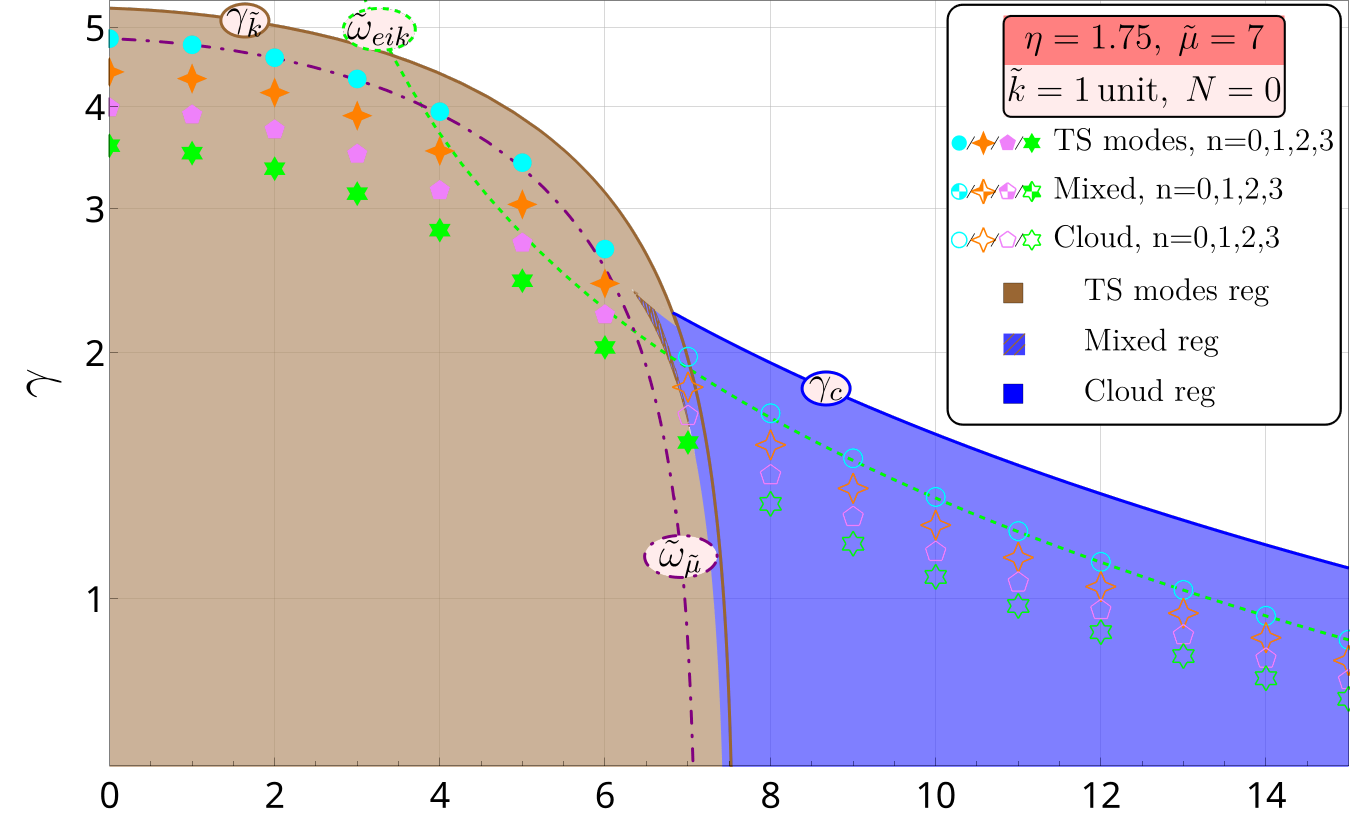}
	}
	\subfigure{\includegraphics[width=0.479\textwidth]{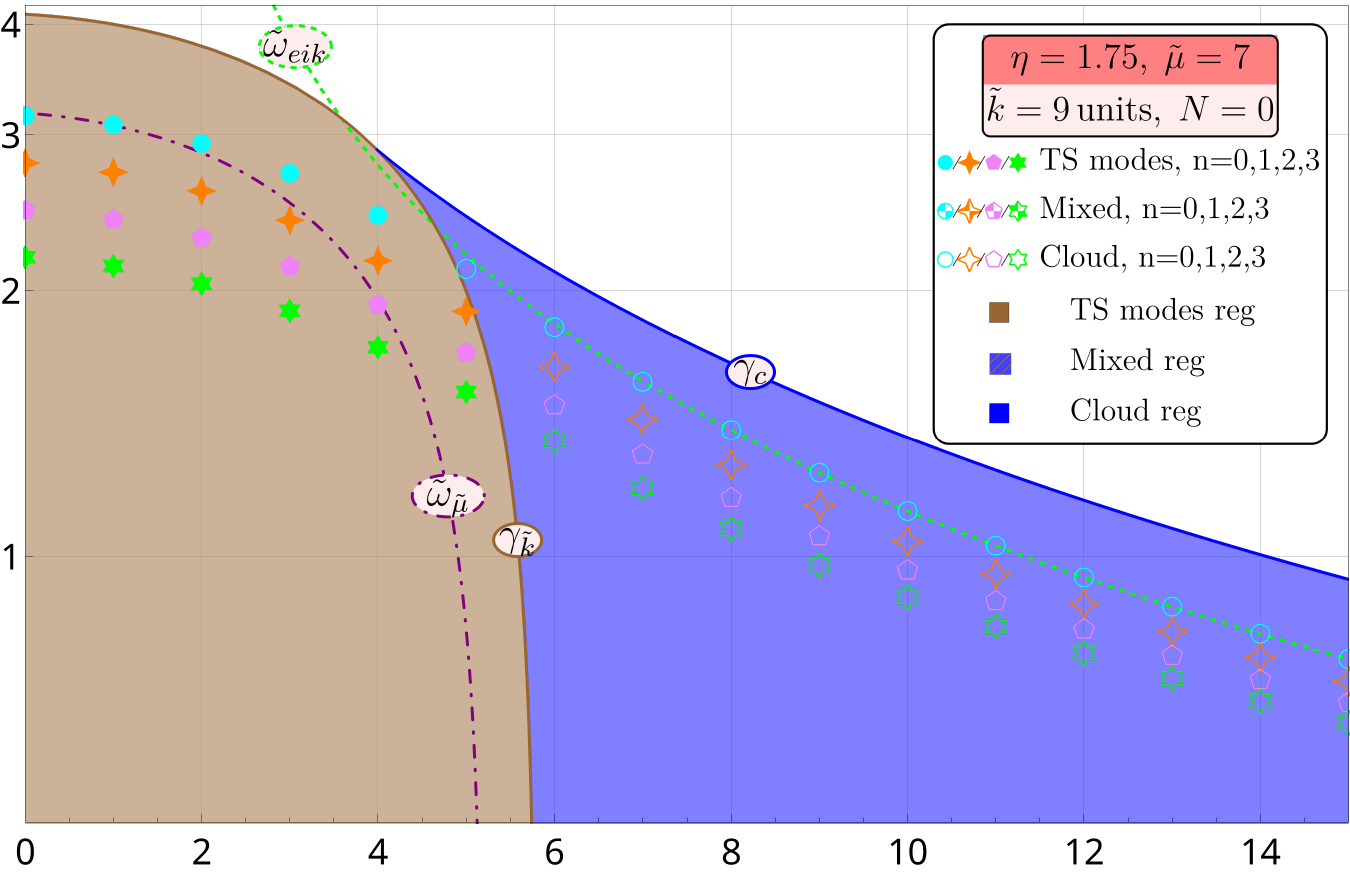}
	}\\
	\subfigure{\includegraphics[width=0.501\textwidth]{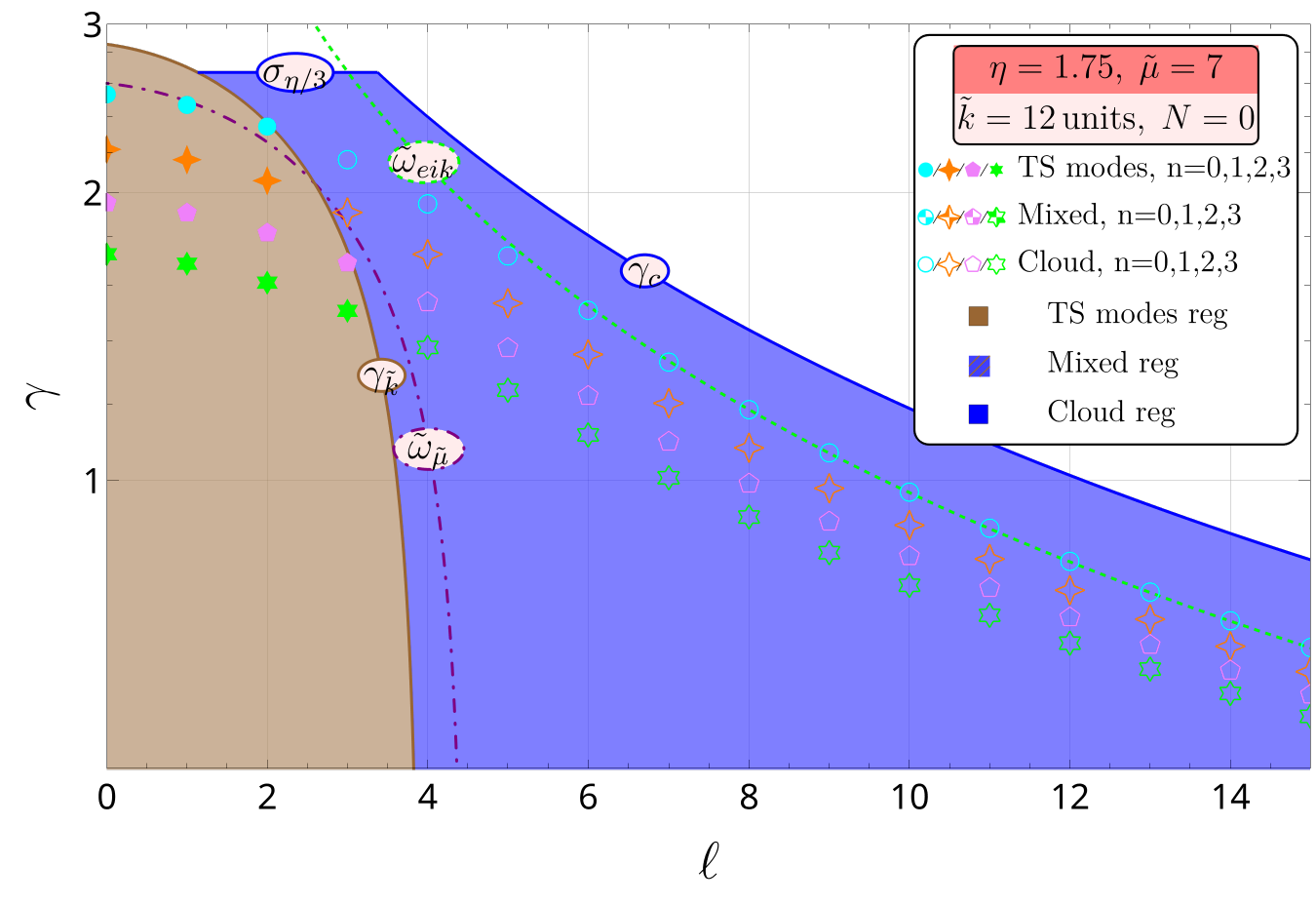}
	}
	\subfigure{\includegraphics[width=0.479\textwidth]{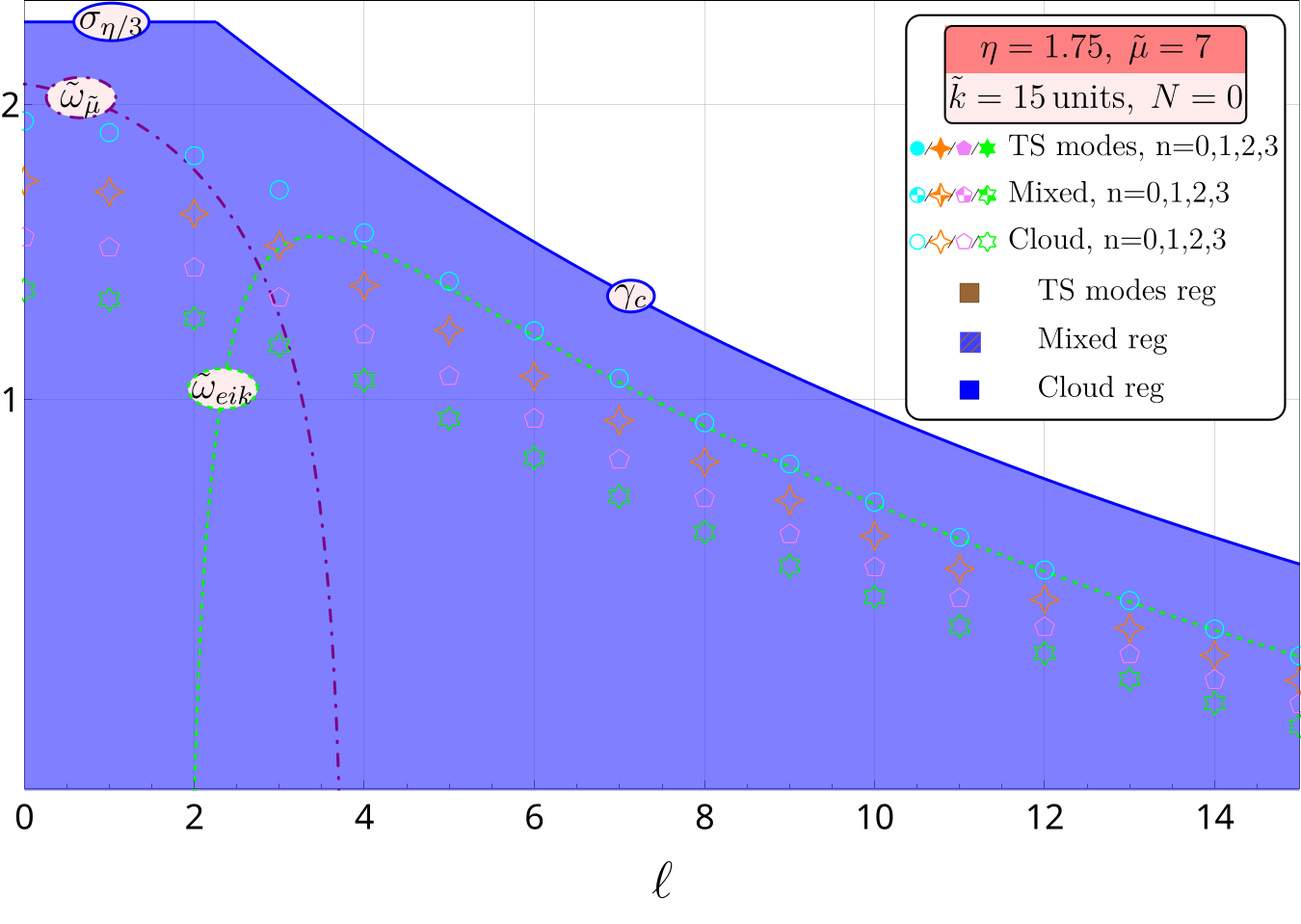}
	}
	\caption{Same as Fig.~\ref{fig:k0mu7eta125Regs}, but for  $\eta=1.75$, $\tilde{\mu}=7$, $N=0$. Now the scalar carries KK momentum $\tilde{k}=1,\,9,\,12,\,15\times\frac{1}{2}\sqrt{\frac{\eta-1}{\eta}}$ (top left, top right, bottom left and bottom right plots, respectively).}
\end{figure*}

This novelty is demonstrated in Fig.~\ref{fig:eta175k191215}, where we have shown the equivalent of Fig.~\ref{fig:k0mu7eta175Regs}, but for non-zero value of KK momentum carried by the scalar. We remind the reader that the latter is quantized, $k=\frac{p}{R_y}$, where $p\in\mathbb{N}$ and $R_y$ is the radius of the extra compact dimension, subject to the regularity condition \eqref{eq:orbifold}. In our dimensionless units (and also setting $R_{y}=1$):
\begin{equation}
	\tilde{k}=\frac{p\,K}{2}\,\sqrt{\frac{\eta-1}{\eta}}.
\end{equation}
Moreover, we have only looked at the case of orbifold parameter $K=1$ in this work, hence we label the KK momentum as given by $p$ units of $\frac{1}{2}\sqrt{\frac{\eta-1}{\eta}}$, and we only indicate the value of $p$ on the plots. For $\eta=1.75$ and $\tilde{\mu}=7$, Eq.~\eqref{eq:noTSCond2} gives us a maximum of $p=24$ units allowed and a transition to a situation with no TS modes for at least $p=15$ units. Indeed, on the bottom right graph of Fig.~\ref{fig:eta175k191215}, where the case of $p=15$ is shown, there is no TS region and we have only found cloud modes. Moreover, at least for these parameters, the mixed region shrinks very quickly. It is barely visible on the top left plot in Fig.~\ref{fig:eta175k191215}, which is otherwise not much different than Fig.~\ref{fig:k0mu7eta175Regs}, and demonstrates the effect of the minimum amount of KK momentum allowed. For $p=9$ and $p=12$ there is no mixed region at all. Of course, once the TS region is gone, so is the mixed one. We do not have an analytical expression that we can refer to in order to determine the conditions for its shrinking. The regions in Fig.~\ref{fig:eta175k191215} were determined numerically.

In addition, one can also see in Fig.~\ref{fig:eta175k191215} how the large scalar mass approximation of Eq.~\eqref{eq:largeMuApprox} loses validity with increasing $\tilde{k}$. This is expected, as Eq.~\eqref{eq:largeMuApprox} assumes that all parameters are $\mathcal{O}(1)$ with respect to $\tilde{\mu}\gg1$, and the plots on the bottom row already have $\tilde{k}\sim\tilde{\mu}$. Furthermore, the approximation~\eqref{eq:largeMuApprox} is designed to capture TS modes, which are localized near the star surface, and thus it should break down once the TS region is gone.

\begin{figure*}[t]
	\label{fig:eta175ell10k5914}
	\centering
	\includegraphics[width=\linewidth]{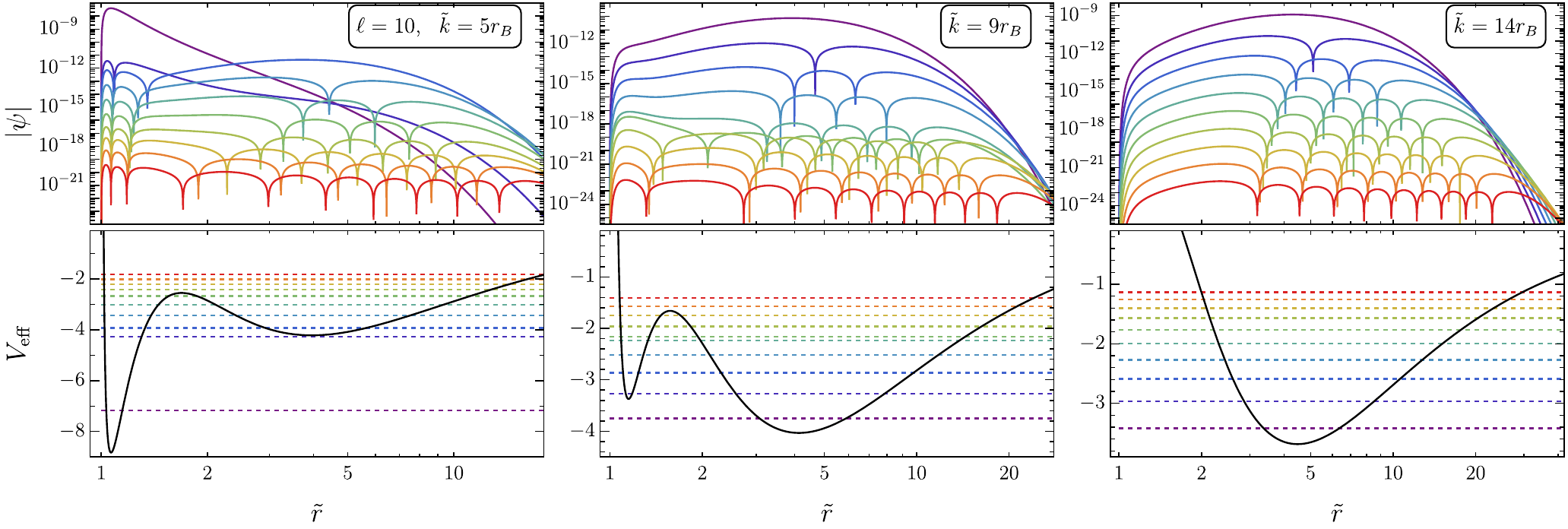}
	\caption{Same as Fig.~\ref{fig:eta125k0wavefunctions}, except we fix $\ell=10$ and look at different values of the KK momentum carried by the field. From left to right, we take $\tilde{k}=(5,\,9,\,14)\times\frac{1}{2}\sqrt{\frac{\eta-1}{\eta}}$, respectively.}
\end{figure*}

Next, we present the equivalent of Fig.~\ref{fig:eta125k0wavefunctions}, but this time we fix $\ell=10$ and vary the amount of the scalar field's KK momentum on the different plots in Fig.~\ref{fig:eta175ell10k5914}. This serves to show how the increase in $\tilde{k}$ slowly pushes all the modes out, until we have only cloud ones left. For $\eta=1.25$ and $\tilde{\mu}=7$, the parameters in Fig.~\ref{fig:eta125k0wavefunctions}, the bounds in Eq.~\eqref{eq:noTSCond2} are: maximum of $p=62$ and no TS region for $p=37$ and higher. For small values of KK momentum, as on the leftmost figure, the qualitative features of the $\tilde{k}=0$ case are still visible: as we increase the number of overtones we transition from TS to mixed and back to TS modes. The first transition does happen earlier, at $n=2$ in comparison to $n=4$ in Fig.~\ref{fig:eta125k0wavefunctions}. Further increasing $\tilde{k}$ (middle plot) turns the initial TS modes into cloud ones (due to the TS region shrinking) with the rest of the transitions as before, until finally on the rightmost plot there are no more TS or mixed modes.

\subsection{Charged modes and gravitating Thomson dipoles}

In this section we discuss the case in which the scalar field is electrically charged. We remind the reader that, for regularity reasons, the field's fundamental charge $e$ and the star's magnetic charge $P$ are related by Dirac's quantization condition $eP=N/2$, with $N$ any integer: see Eq.~\eqref{eq:DiracQuant}. We choose to work in terms of $P$ and $N$ as independent variables (thus fixing $e=N/2P$), so that $N=0$ corresponds to the neutral case, $N=\pm1$ to the 1-monopole star, $N=\pm2$ to the 2-monopole star, etc. The electromagnetic coupling enters the radial equation \eqref{eq:radeq} only through the constant $\Lambda$, given by
\begin{equation}\label{eq:LLLambda}
    \Lambda=\ell\left(\ell+1\right)-\left(N/2\right)^{2}\, ,
\end{equation}
so the mode classification and other analysis about the radial function and the normal frequencies of Sections \ref{sec:ZeroKK} and \ref{sec:KK} also accounts for charged scalars. In certain aspects, charged and neutral modes are almost identical. For example, two normal modes with the same $\tilde{\mu}$ and $\Lambda$ will possess identical radial profiles (up to global normalization) and frequency spectrum, regardless of the values that $N$ and the respective $\ell$'s may take. However, regarding other characteristics they can be strikingly different. Even if a neutral and a charged mode have identical frequencies, radial profiles and total energies, they exhibit distinct spatial distributions due to the difference between Wu-Yang monopole harmonics and ordinary spherical harmonics (see Appendix \ref{A1}). Consequently, such two modes possess different amounts of angular momentum, with the charged field consistently capable of carrying a larger value.

To see this, we need to make a physically meaningful comparison between neutral and charged fields. We first notice that the contributions to the 4-dimensional ADM energy $E[\varphi]$ and $z$-axis angular momentum $L[\varphi]$ due to a general normal mode \eqref{eq:mode} are given (up to numerical prefactors that are unimportant here) by the conserved functionals in \eqref{eq:ConForm} on a constant-$t$ surface $\Sigma_{t}$,
\begin{widetext}
    \begin{align}
     E[\varphi]&=-\int_{\Sigma_{t}}\bold{I}_{\partial_{t}}\left[\varphi\right]=\pi r_{B}\int_{1}^{\infty}d\tilde{r}\left\{\left(\lvert\tilde{\omega}\rvert^{2}\frac{\tilde{r}^{3}}{\tilde{r}-1/\eta}+\tilde{\mu}^{2}\tilde{r}^{2}+\Lambda\right)\lvert\psi\rvert^{2}+(\tilde{r}-1)(\tilde{r}-1/\eta)\lvert d\psi/d\tilde{r}\rvert^{2}\right\}\, , \\ \label{eq:angmodes}
     L[\varphi]&=\int_{\Sigma_{t}}\bold{I}_{\partial_{\phi}}\left[\varphi\right]=2\pi r^{2}_{B}\tilde{\omega} m\int_{1}^{\infty}d\tilde{r}\frac{\tilde{r}^{3}}{\tilde{r}-1/\eta}\lvert\psi\rvert^{2}\,,
    \end{align}
\end{widetext}
where we have set $R_{y}=1$ and restricted to zero-KK modes for simplicity. Consider now two normal modes with equal masses $\tilde{\mu}$, but one being neutral $\varphi_{\text{n}}$ (so $N_{\text{n}}=0$) and the other charged $\varphi_{\text{c}}$ (so $N_{\text{c}}\ne0$) and possessing different angular numbers $\ell_{\text{n}}$ and $\ell_{\text{c}}$, respectively. Furthermore, assume that the charge number $N_{\text{c}}$ can be taken such that $\Lambda_{\text{n}}=\Lambda_{\text{c}}$, where in general $\Lambda$ is given by \eqref{eq:LLLambda} (such $N_{\text{c}}$ does not always exist for arbitrary values of $\ell_{\text{n,c}}$, but it is convenient to restrict to this case in this discussion). Two such modes have the same frequency spectrum, and their radial wavefunctions are also identical if we rescale them suitably to impose that they have the same total energy, $E[\varphi_{\text{n}}]=E[\varphi_{\text{c}}]$. However, recall that the angular numbers must satisfy the conditions below:
\begin{equation}
\begin{aligned}
     \text{Neutral:}&\quad  \ell_{\text{n}}=0,1,2,... \, ,\ \ m_{\text{n}}=-\lvert \ell_{\text{n}} \rvert,-\lvert \ell_{\text{n}} \rvert+1,...,\lvert \ell_{\text{n}} \rvert\, ,\\ \\
   \text{Charged:}&\quad \ell_{\text{c}}=\lvert N_{\text{c}}/2\rvert,\lvert N_{\text{c}}/2\rvert+1,\lvert N_{\text{c}}/2\rvert+2,...\, , \\ 
   &\quad m_{\text{c}}=-\lvert \ell_{\text{c}} \rvert,-\lvert \ell_{\text{c}} \rvert+1,...,\lvert \ell_{\text{c}} \rvert\,.  
\end{aligned}
\end{equation}
From the fact that $\Lambda_{\text{n}}=\Lambda_{\text{c}}$ it is clear that $\ell_{\text{n}}<\ell_{\text{c}}$, so in the charged case $\lvert m_{\text{c}}\rvert$ can take larger values than $\lvert m_{\text{n}}\rvert$. Thus, the charged field can carry more angular momentum, as follows from Eq.~\eqref{eq:angmodes}. In particular, focusing on the field states that possess a maximum amount of angular momentum, $\ell_{\text{n,c}}=m_{\text{n,c}}$, we conclude that
\begin{equation}\label{eq:angularhierarchy}
    \Bigl\lvert L[\varphi_{\text{n}},\ell_{\text{n}}=m_{\text{n}}]\Bigr\rvert< \Bigl\lvert L[\varphi_{\text{c}},\ell_{\text{c}}=m_{\text{c}}] \Bigr\rvert\, .
\end{equation}
To explain this difference, it is useful to illustrate how $\varphi_{\text{n,c}}$ distribute in space. For concreteness, we shall restrict ourselves to fundamental modes with zero KK momentum and take the same background parameters and field mass of Fig.~\ref{fig:k0mu7eta175Regs} ($\eta=1.75$, $\tilde{\mu}=7$). For $\varphi_{\text{n}}$, we shall take maximum angular momentum states with $\ell_{\text{n}}=m_{\text{n}}=6,7,8$, which according to the mode classification given in Section~\ref{sec:ZeroKK} correspond to TS, mixed and cloud modes, respectively. To guarantee that $\Lambda_{\text{n}}=\Lambda_{\text{c}}$, in the charged case we also take maximum angular momentum states with $\ell_{\text{c}}=m_{\text{c}}=N_{\text{c}}/2$ and $N_{\text{c}}=84,112,144$. We emphasize that the condition $m_{\text{c}}=N_{\text{c}}/2$ corresponds to a \textit{north monopole mode}, that is, a mode that is localized along the north component of the $z$-axis and carries angular momentum in the $z$-direction~\cite{Pereniguez:2024fkn} (see also Appendix \ref{A1}). In all cases, we normalize the wavefunctions by setting the total energy to $E[\varphi_{\text{n,c}}]/r_{B}=10^{-3}$. 

\begin{figure*}[!h]
\includegraphics[width=0.85\linewidth]{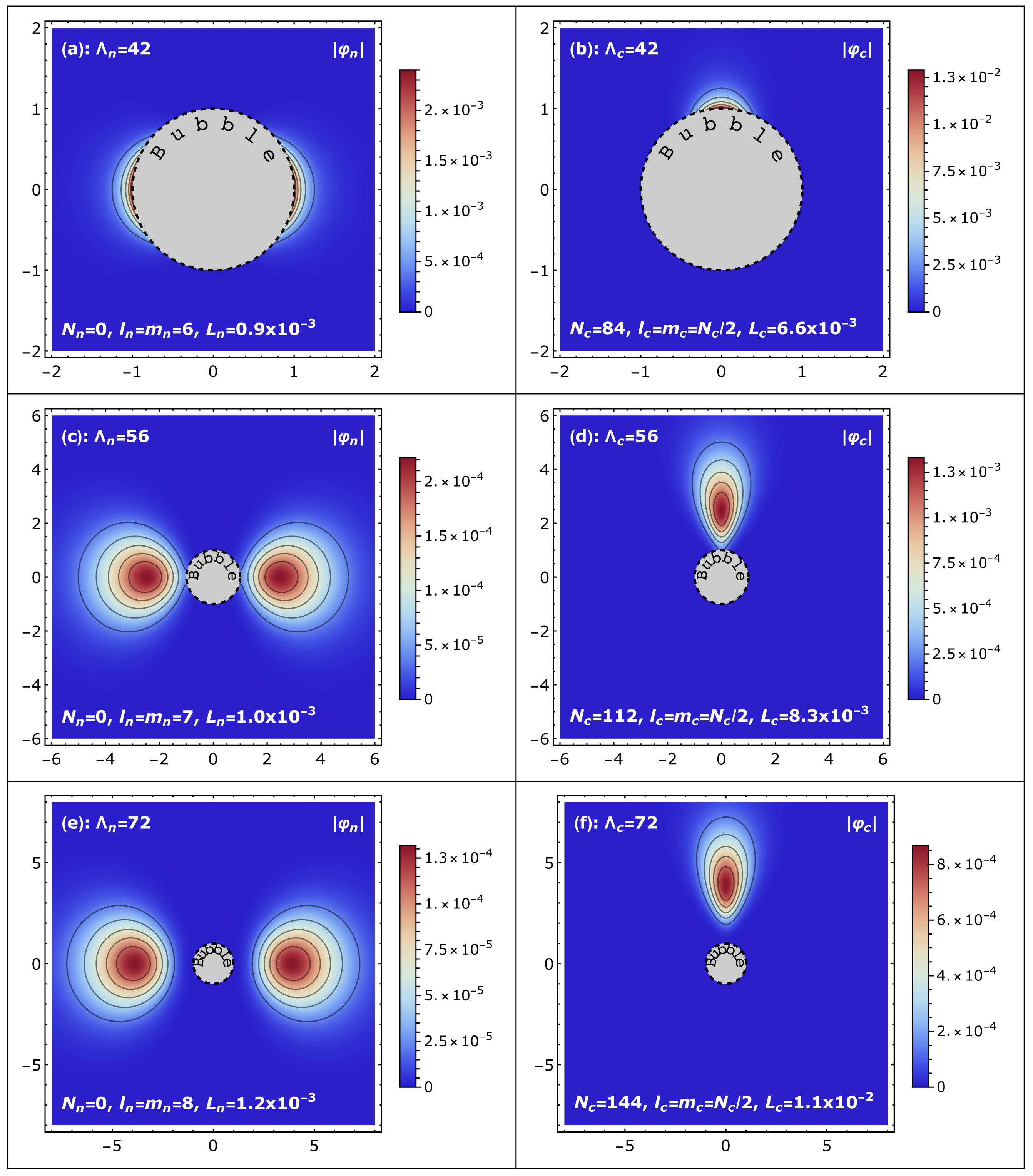}
\caption{Field amplitudes $\lvert \varphi_{\text{n,c}}\rvert$ in the plane spanned by $\tilde{x}\equiv\tilde{r}\sin\theta$ (horizontal axis) and $\tilde{z}\equiv \tilde{r}\cos\theta$ (vertical axis), for a TS background with $\eta=1.75$ and a scalar field mass $\tilde{\mu}=7$ (matching Fig.~\ref{fig:k0mu7eta175Regs}). The bubble's surface (dashed black circle) is at $\tilde{r}=\sqrt{\tilde{x}^{2}+\tilde{y}^{2}}=1$, and the coordinate ranges are adapted to capture the cloud's effective size. As explained in the main text, the wavefunctions are normalized by fixing their total energies to $E[\varphi_{\text{n,c}}]/r_{B}=10^{-3}$, and constant-amplitude contours (solid black lines) are included to guide the eye. Each panel shows the fundamental mode with the indicated values of $N_{\text{n,c}},\ell_{\text{n,c}},m_{\text{n,c}}$, and displays the corresponding values of $\Lambda_{\text{n,c}}$ and total (dimensionless) angular momentum $L_{\text{n,c}}\equiv L[\varphi_{\text{n,c}}]/r_{B}^{2}$. Left and right columns correspond to neutral ($N_{\text{n}}=0$) and charged configurations, respectively. Rows share identical frequencies (which can be obtained from Fig.~\ref{fig:k0mu7eta175Regs}) and radial profiles, and correspond to TS modes (panels a,\,b), mixed modes (panels c,\,d) and cloud modes (panels e,\,f).}
\label{fig:ModesPannels}
\end{figure*}

The results are displayed in Fig.~\ref{fig:ModesPannels}, which we discuss next. As a first observation, we see that in both neutral and charged configurations (left and right columns, respectively) the three types of modes introduced in Section~\ref{sec:ZeroKK} are clearly distinguished: TS modes (top row) are localized near the surface of the bubble, mixed modes (middle row) probe the surface of the bubble but are localized at a larger radius, and cloud modes (bottom row) are localized entirely in a region that does not intersect the bubble. As anticipated earlier, the charged fields for our choice of parameters (in particular $m_{\text{c}}=N_{\text{c}}/2$) correspond to north modes, which exist at the northern $z$-axis but vanish at the southern one (a south mode, corresponding to $m_{\text{c}}=-N_{\text{c}}/2$, would have the same form but upside-down). Given that they are localized in a smaller region of space than the neutral modes but carry the exact same amount of energy by construction, the maximum field amplitude is (roughly ten times) higher in the charged case. Regarding the angular momentum, in agreement with Eq.~\eqref{eq:angularhierarchy} we see that the charged modes carry a larger amount. In the neutral case, the angular momentum is of orbital nature, i.e. due to the rotation of the field around the bubble. The situation is less intuitive for charged fields. Heuristically, the origin of the angular momentum in that case can be explained from the properties of the Thomson dipole~\cite{thomson_2009}. The latter is simply a pair of point charges, one electric and one magnetic, which are separated and at rest. Such a static configuration has a non-trivial Poynting vector, which endows the electromagnetic field with an angular momentum proportional to the product of the charges, and pointing from the electric to the magnetic charge (see~\cite{Garfinkle:1990zx,Bunster:2007sn,Dyson:2023ujk,Pereniguez:2024fkn} for the BH generalization). With this picture in mind, it is clear that the charged field configurations in the second column of Fig.~\ref{fig:ModesPannels} are simply a wave-like version of a Thomson dipole, where the bubble plays the role of the magnetic charge and the field replaces the electric charge. Consequently, these states carry angular momentum along the $z$-direction, even though they are localized along the $z$-axis. 

\section{Discussion}\label{sec:discussion}

We have studied the structure of massive bound states in the vicinity of TSs, focusing on charged scalars minimally coupled to the Einstein--Maxwell theory. First, we have established generic properties expected to hold for other compact objects -- namely, that bound states are strictly normal, and the regimes $r_{b}/\lambda\ll1$ and $r_{b}/\lambda\gg1$ are hydrogen-like and governed by congruences of timelike geodesics, respectively. Next, we have proposed a bound-mode classification based on the classically allowed regions for the modes, and their position relative to the ISCO. In the case of TSs, considering modes with no KK momentum, modes fall in three categories: TS modes that exist close to the star's center, cloud modes that exist beyond the ISCO, and mixed modes that experience two classically allowed regions separated by a centrifugal barrier. If KK momentum is included, additional potential barriers appear and the mode classification becomes richer. In particular, we have found that modes with too large KK momentum cannot remain bound to the star, and can only be radiative. In the case that the field is electrically charged, we have shown that, while the spectrum and radial distribution resemble closely (or are even identical to) the neutral case, their spatial configuration changes drastically. In fact, we have shown that TSs and charged-field bound states provide a smooth horizonless realization of a gravitating Thomson dipole, so far only constructed with BHs. 

This work shows that TSs yield genuine realizations of a gravitational atom, superseding BH ones in that they are truly stationary, linearly stable systems. However, unlike for BHs where clouds grow spontaneously due to superradiant instabilities, it is unclear how TS gravitational atoms may form. One possibility consists in assuming a non-zero scalar-field background density, motivated from dark matter considerations~\cite{Hui:2019aqm,Clough:2019jpm}, so that the massive states described above would unavoidably form around TSs. 

A mechanism with less assumptions would be that TS gravitational atoms form spontaneously, and correspond to end states of instabilities, in closer analogy to the BH case. However, it is unlikely that the instability is of superradiant nature -- a fact that is in agreement with the intuition that TSs correspond to microscopic fundamental states. The reason is that in the absence of an horizon there is no dissipation channel in the scattering of waves, which is one of the fundamental ingredients of superradiance~\cite{1972JETP...35.1085Z,Bekenstein:1998nt}. This is for instance the case of ideal fluid stars, which do not experience superradiance, and dissipative effects such as conductivity~\cite{Richartz:2013unq,Cardoso:2017kgn} or viscosity~\cite{Redondo-Yuste:2025ktt} need to be accounted for in order to trigger superradiance. The absence of superradiance in the spherical TSs considered here follows easily: massless modes at infinity behave as $\psi(r)\sim e^{-i \omega r}+R e^{i \omega r}$, corresponding to incoming and outgoing waves, where $R$ is the reflectivity coefficient. If the frequency is real, then the regularity condition \eqref{eq:rb} implies that $\psi(r)$ and $\bar{\psi}(r)$ are linearly \textit{dependent} solutions, and by the vanishing of the Wronskian one finds $\lvert R \rvert=1$, corresponding to perfect reflectivity -- so, in particular, there is no amplification. A similar argument is expected to hold for rotating TSs, only recently constructed in~\cite{Bianchi:2025uis,Heidmann:2025pbb}, as long as a regularity condition similar to Eq.~\eqref{eq:rb} holds at the star's center. However, it would be interesting to explore whether the electromagnetic coupling can change this picture, as it has been shown to yield new regions of negative-energy states and different phenomenology in the BH case~\cite{Dyson:2023ujk,Pereniguez:2024fkn}.

Alternatively, in Ref.~\cite{Heidmann:2025pbb} the authors notice that the rotating solution presents a 5-dimensional ergoregion. This indicates that KK-modes might be subject to a different class of instability not based on superradiant scattering, known as ergoregion instability~\cite{Friedman1978} (see e.g.~\cite{Cardoso:2005gj} in the context of microstate geometries). This is associated to spaces with ergoregions, but no horizons, where a field grows unbounded as long as it is in a negative energy state and it radiates to infinity. We have shown that massive modes with large-enough KK momentum are \textit{necessarily} radiative. Assuming that this is also true for the rotating TS, combined with the 5-dimensional ergoregion of Ref.~\cite{Heidmann:2025pbb}, it follows that massive modes with enough KK momentum can be radiative and simultaneously occupy negative-energy states, thus meeting the assumptions that lead to an ergoregion instability. Understanding whether this instability can yield, as an end state, a TS with a stationary bound-state cloud such as those considered here requires tracking non-linearly the evolution of the instability. The knowledge of explicit rotating TS solutions~\cite{Bianchi:2025uis,Heidmann:2025pbb} allows a precise analysis of these questions, at least at the perturbative level, where one can sharply test the expectations discussed here. 

\acknowledgements

We thank Pierre Heidmann, Paolo Pani and Jaime Redondo-Yuste for insightful discussions. D.P. acknowledges financial support by the VILLUM Foundation (grant
no. VIL37766) and the DNRF Chair program (grant
no. DNRF162) by the Danish National Research Foundation. The Center of Gravity is a Center of Excellence funded by the Danish National Research Foundation under grant No. 184. This project has received funding from the European
Union’s Horizon 2020 research and innovation programme
under the Marie Sklodowska-Curie grant agreement No
101007855 and No 101007855. 
E.B., D.P. and N.S. are supported by NSF Grants No.~AST-2307146, No.~PHY-2513337, No.~PHY-090003, and No.~PHY-20043, by NASA Grant No.~21-ATP21-0010, by John Templeton Foundation Grant No.~62840, by the Simons Foundation [MPS-SIP-00001698, E.B.], by the Simons Foundation International [SFI-MPS-BH-00012593-02], and by Italian Ministry of Foreign Affairs and International Cooperation Grant No.~PGR01167.
This work was carried out at the Advanced Research Computing at Hopkins (ARCH) core facility (\url{https://www.arch.jhu.edu/}), which is supported by the NSF Grant No.~OAC-1920103. IB, BG are supported
in part by the Simons Collaboration on Global Categorical Symmetries and also by the NSF grant PHY-
2412361. 

\appendix

\section{Wu-Yang monopole harmonics}\label{A1}
	
	The Wu-Yang monopole harmonics $Y_{N,\ell,m}$~\cite{Wu:1976ge} are eigenfunctions of a well-defined set of angular momentum operators, in the background of a magnetic monopole. They live in a $U(1)$-line bundle, in the sense that under a gauge-transformation \eqref{eq:gauge} they transform as $Y_{N,\ell,m}\mapsto e^{i\alpha}Y_{N,\ell,m}$. This point is reviewed explicitly in~\cite{Pereniguez:2024fkn}, and here we simply write the resulting explicit expressions for $Y_{N,\ell,m}$:
	\begin{equation}
		\begin{aligned}\label{Yexp}
			 Y_{N,\ell,m}(\theta,\phi)&=\mathcal{P}_{N,\ell,m}(\theta)e^{i m\phi}\\
            &=\mathcal{N} (1-x)^{\frac{\vert \alpha \vert}{2}}(1+x)^{\frac{\vert \beta \vert}{2}} P^{(\vert \alpha \vert,\vert \beta \vert)}_{\nu}(x)e^{i m\phi}\, , \\ \\
			 \ell=\lvert N\rvert/2 , \lvert& N \rvert/2+1, ... \, ,  \ \ m=-\ell,-\ell+1,...,\ell\, ,
		\end{aligned}
	\end{equation}
	where $x=\cos\theta$, $P^{(a,b)}_{n}(x)$ denote the usual Jacobi polynomials, the various constants are
	\begin{align}
		\alpha&=N/2-m\, ,\ \ \ \beta=-N/2-m\, , \\  
        \nu&=\ell +m+\frac{\alpha-\lvert \alpha \rvert+\beta-\lvert \beta \rvert}{2} \, ,\\ 
        \mathcal{N}&=\frac{(-1)^{\frac{\alpha-\lvert \alpha \rvert}{2}}}{\sqrt{4\pi}}\sqrt{\frac{2\ell+1}{ 2^{\lvert \alpha \rvert+\lvert \beta \rvert}}\frac{\nu!(\nu+\lvert \alpha \rvert+\lvert \beta \rvert)!}{(\nu+\lvert \alpha \rvert)!(\nu+\lvert \beta \rvert)!}}\, ,
	\end{align}
	and $\mathcal{N}$ has been fixed so that the usual orthogonality condition holds,
	\begin{equation}
		\int_{S^{2}}\bar{Y}_{N,\ell',m'}Y_{N,\ell,m}d\Omega = \delta_{\ell'\ell}\delta_{m'm}\, .
	\end{equation}
It is useful to note the characteristic problem that the $Y_{N,\ell,m}$'s solve. Consider the Euclidean 3-dimensional space, a monopole field $A=-P\cos\theta d\phi$, and introduce the gauge-covariant operators,
	\begin{equation}\label{eq:angmom}
		\bold{D}=\boldsymbol{\nabla}+i e A\, ,\quad L_{l}=-i\epsilon_{ljk}x^{j}\bold{D}^{k}+\frac{N}{2}\frac{x^{l}}{r}\, ,
	\end{equation}
where $\boldsymbol{\nabla}$ denotes the Euclidean covariant derivative. Then $Y_{N,\ell,m}$ satisfy
\begin{align}
    L^{2}Y_{N,\ell,m}&=\ell(\ell+1)Y_{N,\ell,m}\, ,\\ 
    \bold{D}^{2}Y_{N,\ell,m}&=-\left[\ell(\ell+1)-\left(N/2\right)^{2}\right]Y_{N,\ell,m}\, .
\end{align}
Table~\ref{monopoleharms} shows the explicit form of some low-order monopole harmonics, in the gauge considered here ($A=-P\cos\theta d\phi$). 
\begin{widetext}
    \center
	\begin{table}[h!]
		\begin{equation*}
			\begin{array}{l!{\vline width 1pt}c|c|c}
				\sqrt{4\pi}Y_{N,\ell,m} & q=0 & N=1 & N=2  \\ 
				\Xhline{1pt}
				\ell=0 & 1 & - & -  \\ \hline
				\ell=1/2 & - & \mp \sqrt{1\mp x} e^{\pm i \frac{1}{2}\phi} & -  \\ \hline
				\ell=1 & \begin{aligned} \sqrt{3} x \\ \mp\sqrt{3/2}\sqrt{1-x^{2}}e^{\pm i \phi} \end{aligned} & - & \begin{aligned} -\sqrt{3/2}\sqrt{1-x^{2}} \\ \sqrt{3/4}(1\mp x)e^{\pm i \phi} \end{aligned} \\ \hline
				\ell=3/2 & - & \begin{aligned} -\sqrt{1/2}\sqrt{1\mp x}(1\pm 3 x)e^{\pm i\frac{1}{2}\phi} \\ \sqrt{3/2}\sqrt{1\pm x}(1\mp x)e^{\pm i\frac{3}{2}\phi} \end{aligned} & - \\ \Xhline{1pt}
			\end{array}
		\end{equation*}
		\caption{Some low-order monopole harmonics.}
		\label{monopoleharms}
	\end{table}
\end{widetext}
	It is worth noticing the special harmonics with $m=N/2$ ($m=-N/2$), the \textit{north (south) monopole harmonics} introduced in~\cite{Pereniguez:2024fkn}. They are non-vanishing at the north (south) pole $\cos\theta=+1$ ($\cos\theta=-1$), but vanish at the south (north) one $\cos\theta=-1$ ($\cos\theta=+1$). These have no analogue in neutral or electrically charged backgrounds, and hence lead to distinctive properties of magnetically charged objects. 

\section{Equations of Motion and Bianchi Identities}\label{A2}
	
	The equations of motion derived from \eqref{eq:theory} read
	\begin{equation}
		\begin{aligned}
			&G_{\mu\nu}=T^{F}_{\mu\nu}+T^{\Phi}_{\mu\nu}\, , \quad \nabla^{\mu} F_{\mu\nu}=-e J^{\Phi}_{\nu}\, , \quad E^{\Phi}=0\, ,
		\end{aligned}
	\end{equation}
	where
	\begin{equation}\label{eq:emTensA2}
		\begin{aligned}
			&T^{F}_{\mu\nu}= F_{\mu}^{\ \alpha}F_{\nu\alpha}-\frac{1}{4}F^{2}g_{\mu\nu}\,, \\
            &T^{\Phi}_{\mu\nu}=D_{(\mu}\bar{\Phi}D_{\nu)}\Phi-\frac{1}{2}\left(D_{\alpha}\bar{\Phi} D^{\alpha}\Phi+\mu^{2} \bar{\Phi}\Phi\right)g_{\mu\nu}\,, \\
			&J^{\Phi}_{\mu}=\frac{i}{2}\left(\bar{\Phi}D_{\mu}\Phi-\Phi D_{\mu}\bar{\Phi}\right)\, ,\\ &E^{\Phi}=D^{\mu}D_{\mu}\Phi-\mu^{2}\Phi\, ,
		\end{aligned} 
	\end{equation}
	and one can show that the following identities hold:
	\begin{equation}
		\begin{aligned}
			&\star d \star J^{\Phi}=-\frac{i}{2}\left(\bar{\Phi} E^{\Phi}-c.c.\right)\, , \\  &\nabla^{\mu}T^{F}_{\mu\nu}=F_{\nu\alpha}\nabla^{\mu}F_{\mu}^{\ \alpha}\, ,\\
            &\nabla^{\mu}T^{\Phi}_{\mu\nu}=\frac{1}{2}\left(E^{\Phi}D_{\nu}\bar{\Phi}+c.c.\right)+e F_{\nu}^{\ \mu}J^{\Phi}_{\mu}\,.
		\end{aligned}
	\end{equation}
	These may be directly verified, or derived as the Noether identities following from diffeomorphism and gauge invariance of the action \eqref{eq:theory} (see e.g.~\cite{Ortin:2022uxa,Pereniguez:2024fkn,Dyson:2023ujk}).
	
	\section{Quasimodes in the eikonal limit}\label{secA:quasimodes}
	
	Quasimodes are approximate solutions to the wave equation localized at the minimum of the potential. They allow us to determine the real part of the frequency in an appropriate limit. As we construct them here, they are insensitive to the imaginary part of the frequency, that vanishes for bounded modes on a TS background (as shown in the main text). However, the same method can also be applied to the black string regime, which we do not show here, but we have checked that the method works well. In order to simplify the algebra, we use the dimensionless coordinates of Eq.~\eqref{eq:dimlessVars}, together with the following radial variable:
	\begin{equation}\label{eq:varOnMin}
		\tilde{r}=\frac{2\,\eta\,\ell\,(\ell+1)\,z}{\big(\tilde{k}^2\,(1-\eta)+\tilde{\mu}^2\big)}+1,
	\end{equation}
	The equation becomes:
	\begin{equation}\label{eq:eqQuasi}
		g(z)\,\partial_z\,\big[g(z)\,\psi'(z)\big]-V(z)\,\psi(z)=0,
	\end{equation}
	where
	\begin{gather}
		g(z)=z\,\Bigg(1-\frac{2\,\eta^2\,\ell\,(\ell+1)\,z}{(1-\eta)\big(\tilde{k}^2\,(1-\eta)+\tilde{\mu}^2\big)}\Bigg),\notag\\
		V(z)=-\Big(\frac{\eta}{1-\eta}\,\tilde{k}^2+v_1\,z+v_2\,z^2+v_3\,z^3+v_4\,z^4\Big),
	\end{gather}
	with
	\begin{widetext}
	\begin{gather}
		v_1=\frac{2\,\eta^2\,\ell\,(\ell+1)\big(\tilde{k}^2\,(3-4\,\eta)-(1-\eta)\,\Lambda+(1-\eta)\,\tilde{\mu}^2+\eta\,\tilde{\omega}^2\big)}{(1-\eta)^2\,\big(\tilde{k}^2\,(1-\eta)+\tilde{\mu}^2\big)},\notag\\
		v_2=\frac{4\,\eta^3\,\ell^2\,(\ell+1)^2\,\big(\tilde{k}^2\,(3-6\,\eta)+2\,\tilde{\mu}^2+\eta\,(\Lambda-3\,(\tilde{\mu}^2-\tilde{\omega}^2))\big)}{(1-\eta)^2\,\big(\tilde{k}^2\,(1-\eta)+\tilde{\mu}^2\big)^2},\notag\\
		v_3=\frac{8\,\eta^4\,\ell^3\,(\ell+1)^3\big(\tilde{k}^2\,(1-4\,\eta)+(1-3\,\eta)\,\tilde{\mu}^2+3\,\eta\,\tilde{\omega}^2\big)}{(1-\eta)^2\,\big(\tilde{k}^2\,(1-\eta)+\tilde{\mu}^2\big)^3},\notag\\
		v_4=\frac{16\,\eta^6\,\ell^4\,(\ell+1)^4\,\big(\tilde{\omega}^2-(\tilde{k}^2+\tilde{\mu}^2)\big)}{(1-\eta)^2\,\big(\tilde{k}^2\,(1-\eta)+\tilde{\mu}^2\big)^4},
	\end{gather}
	\end{widetext}
	and $z=0$ for $\tilde{r}=1$, while $\tilde{r}\to\infty$ corresponds to $z\to\infty$.
	
	Next, we want to identify the minimum we are interested in. To that end we look at $V'(z)=0$ with $\tilde{\omega}=\sum_{j=0}^{\infty}w_j\,\ell^{-j}$, $z=\sum_{j=0}^{\infty}z_j\,\ell^{-j}$ (ignoring the imaginary part of the frequency) in the eikonal limit. We find three extrema and derive along the way $w_0=\sqrt{\tilde{k}^2+\tilde{\mu}^2}$ and $w_1=0$.
	
	First, there is a local minimum near $z_0=0$, but we need to determine higher order terms in its eikonal expansion in order to talk about its position relative to the star surface. To that end, we need to look for the minimum while simultaneously solving the radial equation so as to obtain higher order $w_j$ pieces. That is rather tedious and we just state the result -- namely, this extremum does not sit outside of the star surface/event horizon when corrections are included. Thus, we will not focus on it here.
	
	Second, there is a local maximum sitting in-between the other two extrema, which is not where we expect the modes to localize.
	
	Finally, we have a local minimum for the outermost extremum. This is the one we are interested in. The choice of prefactor in \eqref{eq:varOnMin} was made intentionally, so that our quasimode construction is in the vicinity of $z=1+\mathcal{O}(\ell^{-1})$ for $\ell\gg1$. 
	
	The order of the just mentioned extrema requires knowledge of $w_2$, but at leading order in the radial equation expansion we get enough information to determine $w_2<0$ (as also required for a bounded mode), which enables the discussion above and thus the beginning of our approximate construction. Furthermore, we also find that for $\sigma<0$, all the minima are inside the TS, confirming our earlier analysis, resulting in the necessary condition for the existence of bound modes in the TS background, $\sigma>0$.
	
	To solve Eq.~\eqref{eq:eqQuasi} in the eikonal regime, we take:
	\begin{equation}\label{eq:approxRadSol}
		\psi(z)=e^{-\ell\,\phi(z)}\,Z(z),
	\end{equation}
	and expand in powers of $\ell$
	\begin{gather}
		Z(z)=Z_0(z)\Bigg[1+\sum_{j=1}^{\infty}Z_j\,\ell^{-j}\Bigg],\quad\tilde{\omega}=\sum_{j=0}^{\infty}w_j\,\ell^{-j},\notag\\ w_0=\sqrt{\tilde{k}^2+\tilde{\mu}^2},\quad w_1=0.
	\end{gather}
	We can then determine $Z_j$ and $w_j$ order by order in $\ell$. At leading order we obtain:
	\begin{equation}
		\phi'(z)^2=\frac{1}{z^2}-\frac{2}{z}-\frac{8\,w_2\,\eta^2\sqrt{\tilde{k}^2+\tilde{\omega}^2}}{\big(\tilde{k}^2(1-\eta)+\tilde{\mu}^2\big)^2}.
	\end{equation}
	We can integrate for $\phi(z)$ and smoothness of the solution over the allowed range for $z$ gives us:
	\begin{equation}
		\phi(z)=z-\log(z)+c_\phi,\quad w_2=-\frac{\big(\tilde{k}^2\,(\eta-1)-\tilde{\mu}^2\big)^2}{8\,\eta^2\,\sqrt{\tilde{k}^2+\tilde{\mu}^2}},
	\end{equation}
	where $c_\phi$ is a constant related to the amplitude of the radial function. We have made a choice of sign for $\phi(z)$, so that $e^{-\ell\,\phi(z)}$ decays away from $z=1$. (It is here that one can determine the sign of $w_2$, before the location of the potential minimum we are interested in has been clarified. It is only after that we can decide which $\phi(z)$ solution to pick.) From here onward, we can keep integrating the resulting equation at every order, looking for a smooth solution, in order to determine $w_j$ and $Z_j$. In particular, we get
	\begin{widetext}
	\begin{align}
		&w_3=\frac{\big(\tilde{k}^2\,(\eta-1)-\tilde{\mu}^2\big)^2}{4\,\eta^2\,\sqrt{\tilde{k}^2+\tilde{\mu}^2}},\notag\\
		w_4=\frac{\big(\tilde{k}^2(1-\eta)+\tilde{\mu}^2\big)^2}{128\,\eta^4\,\big(\tilde{k}^2+\tilde{\mu}^2\big)^{3/2}}\bigg[&\eta^2\,\Big(15\,\tilde{k}^4-4\,\tilde{\mu}^2\,\big(N^2+12\big)-4\,\tilde{k}^2\,\big(N^2-4\,\tilde{\mu}^2+12\big)\Big)\notag\\
		&-6\,\eta\,\tilde{k}^2\,\big(\tilde{k}^2+\tilde{\mu}^2\big)-9\,\big(\tilde{k}^2+\tilde{\mu}^2\big)^2\bigg],\notag\\
		Z_1(z)=\frac{\tilde{k}^2\,(\eta-1)\,(1-\eta+z^2\,(1+2\,\eta))}{4\,\eta^2\,z}&-\frac{(1-\eta+z^2)\,\tilde{\mu}^2}{4\,\eta^2\,z}-\frac{N^2\,z}{8}+c_{Z_1}\notag\\
		&+\Bigg[\frac{3\,\tilde{k}^2}{4}\Bigg(1-\frac{1}{\eta^2}\Bigg)-\frac{(3+\eta)\,\tilde{\mu}^2}{4\,\eta^2}-\frac{N^2}{8}\Bigg]\,\log(z),
	\end{align}
	\end{widetext}
	where $c_{Z_1}$ is an integration constant. The $\log(z)$ term is not a problem, as we have a $z^\ell$ coming from $e^{-\ell\,\phi(z)}$ and by definition our eikonal approximation works only for $\ell\geq1$, which also means that $1/z$ terms are not an issue. Moreover, $Z_0(z)$ turns out to be just a constant that we can absorb in $c_\phi$.
	
	Collecting all the pieces, the expression for the frequency in the eikonal limit is given by:
	\begin{widetext}
	\begin{align}\label{eq:eikonalApprox}
		\tilde{\omega}_{\rm eik}=\sqrt{\tilde{k}^2+\tilde{\mu}^2}-\frac{\sigma^2}{8\,\eta^2\,\sqrt{\tilde{k}^2+\tilde{\mu}^2}}\frac{1}{\ell^2}&+\frac{\sigma^2}{4\,\eta^2\,\sqrt{\tilde{k}^2+\tilde{\mu}^2}}\frac{1}{\ell^3}\notag\\
		+\frac{\sigma^2}{128\,\eta^4\,\big(\tilde{k}^2+\tilde{\mu}^2\big)^{3/2}}\bigg[&\eta^2\,\Big(15\,\tilde{k}^4-4\,\tilde{\mu}^2\,\big(N^2+12\big)-4\,\tilde{k}^2\,\big(N^2-4\,\tilde{\mu}^2+12\big)\Big)\notag\\
		&-6\,\eta\,\tilde{k}^2\,\big(\tilde{k}^2+\tilde{\mu}^2\big)-9\,\big(\tilde{k}^2+\tilde{\mu}^2\big)^2\bigg]\,\frac{1}{\ell^4}+\mathcal{O}\big(\ell^{-5}\big),
	\end{align}
	\end{widetext}
	With this expression to fourth order we can also expand $\gamma=\sqrt{\tilde{k}+\tilde{\mu}^2-\tilde{\omega}^2}$ for $\ell\gg1$:
	\begin{widetext}
	\begin{equation}
		\gamma=\frac{\tilde{k}^2\,(1-\eta)+\tilde{\mu}^2}{2\,\eta\,\ell}-\frac{\tilde{k}^2\,(1-\eta)+\tilde{\mu}^2}{2\,\eta\,\ell^2}+\frac{\big(\tilde{k}^2\,(1-\eta)+\tilde{\mu}^2\big)\,\big(2\,\tilde{k}^2\,(1+\eta-2\,\eta^2)+2\,\tilde{\mu}^2+(8+N^2)\,\eta^2\big)}{16\,\eta^3\,\ell^3}+\mathcal{O}\big(\ell^{-4}\big).
	\end{equation}
	\end{widetext}
	
	Inverting Eq.~\eqref{eq:varOnMin}, we can immediately see that the location of the minimum, where the clouds get localized, and where we expect their peak to be, scales as $\tilde{r}\sim\ell^2$ for $\ell\gg1$. We can also get an estimate for the cloud's size in the eikonal regime by looking at the second derivative of the approximate radial wavefunction~\eqref{eq:approxRadSol} in terms of $\tilde{r}$. The outcome is:
	\begin{equation}
		\frac{4\,\eta\,\sqrt{\ell}\,(\ell+1)}{\big(\tilde{k}^2\,(1-\eta)+\tilde{\mu}^2\big)}.
	\end{equation}
	
	\section{Quasimodes in the large scalar mass limit}\label{secA:quasimodes2}
	The computation in this section is valid for the TS case only, as it targets modes localized in the classically allowed region that extends from the star surface outwards in the language of Section~\ref{sec:ZeroKK} -- that is, the TS modes.
	
	One should be able to replicate the construction from the previous section, but targeting the TS modes, as qualitatively they are also bound states sharply peaked in a local minimum. From the regime of validity of the eikonal approximation (all arguments in both approximations hold for $\tilde{k}=0$, where it is much easier to infer the relevant regime), one anticipates that the appropriate regime to characterize these modes is one where $\tilde{\mu}$ is much larger than any other scale in the problem -- most likely, a regime where the modes are very close to the star's surface.
	
	The mathematical procedure is the same as in the previous section, just in a different limit, namely, $\tilde{\mu}\gg1$. We first look at the derivative of the potential in order to identify a change of coordinates, so that the minimum is localized at $z=1$ to leading order in $\tilde{\mu}\gg1$. The required transformation is:
	\begin{equation}
		\tilde{r}=1+\frac{z}{\tilde{\mu}}\,\Big(\frac{\eta-1}{2\,\sqrt{\eta}}+\tilde{k}\,\sqrt{\eta-1}\Big).
	\end{equation}
	We then proceed with the expansion of the radial equation, defining:	
	\begin{equation}\label{eq:approx2RadSol}
		\psi(z)=z^{\tilde{k}\,\sqrt{\frac{\eta}{\eta-1}}}\,e^{-\hat{\phi}(z)}\,\hat{Z}(z),
	\end{equation}
	where the hats are in order to distinguish this approximation from the one in the previous section and we have prefactored the term we need for the boundary condition at the surface of the star, as the solutions we are looking for do indeed turn out to be very close to it. We then expand in powers of $\tilde{\mu}$:
	\begin{gather}
		\hat{Z}(z)=1+\sum_{j=1}^{\infty}\hat{Z}_j\,\tilde{\mu}^{-j},\quad\tilde{\omega}=\sum_{j=0}^{\infty}w_j\,\tilde{\mu}^{-j},\notag\\ w_0=\tilde{\mu}\,\sqrt{\frac{\eta-1}{\eta}},
	\end{gather}
	where as in the eikonal approximation, $\hat{Z}_0$ is just a constant that can be reabsorbed in $\hat{\phi}(z)$, and $\tilde{\omega}_0$ is what we get from the determination of the leading-order term in the location of the minimum near the star surface. The end result is:
	\begin{widetext}
	\begin{align}\label{eq:largeMuApprox}
		\tilde{\omega}_{\tilde{\mu}}=\tilde{\mu}\,\sqrt{\frac{\eta-1}{\eta}}&+\Bigg(\frac{\sqrt{\eta-1}}{2\,\eta}+\frac{\tilde{k}}{\sqrt{\eta}}\Bigg)-\frac{\sqrt{\eta-1}\,\big(-4\,\eta\,\Lambda-1+8\,\tilde{k}\,\sqrt{\eta(\eta-1)}\big)}{8\,\eta^{3/2}\,\tilde{\mu}}\notag\\
		&+\Bigg(-\frac{(\eta-1)^{3/2}\Big(\eta-1-2\,\eta\,\tilde{k}^2+\tilde{k}\,\sqrt{\eta\,(\eta-1)}\Big)}{4\,\eta^2}-\frac{\sqrt{\eta-1}\,\big(\eta-1+2\,\tilde{k}\,\sqrt{\eta\,(\eta-1)}\big)\,\Lambda}{2\,\eta}\Bigg)\frac{1}{\tilde{\mu}^2},
	\end{align}
	\end{widetext}
	which when plugged in $\gamma=\sqrt{\tilde{k}^2+\tilde{\mu}^2-\tilde{\omega}^2}$ leads to:
	\begin{widetext}
	\begin{equation}
		\gamma=\frac{\tilde{\mu}}{\sqrt{\eta}}-\frac{\eta-1+2\,\tilde{k}\,\sqrt{(\eta-1)\,\eta}}{2\,\eta}+\frac{1+\eta\,\big(4\,(1-\eta)\,\Lambda-\eta\big)+4\,\tilde{k}\,\sqrt{\eta\,(\eta-1)}\,\big(\sqrt{\eta}-2\big)}{8\,\eta^{3/2}\,\tilde{\mu}}+\mathcal{O}\big(\tilde{\mu}^{-2}\big).
	\end{equation}
	\end{widetext}

    \section{Leaver's method}\label{A4}
	Here, we provide more detail about the Leaver method, which we have employed throughout the paper to find numerical mode solutions. Beginning with Eq.~\eqref{eq:radeq} written in dimensionless variables, we take an ansatz for $\psi(\tilde{r})$ which matches the boundary conditions \eqref{eq:rb} and \eqref{eq:rinf}:
	
	\begin{equation}
		\label{eq:Leaver_ansatz}
		\psi(\tilde{r})=e^{\tilde{r}\gamma}\left(\tfrac{\tilde{r}-1}{\tilde{r}-1/\eta} \right)^{\lambda_1}\left(\tilde{r}-\tfrac{1}{\eta} \right)^{\lambda_2}\sum_{n=0}^{\infty}a_n\left(\tfrac{\tilde{r}-1}{\tilde{r}-1/\eta}\right)^n,        
	\end{equation}
	with $\lambda_1=\tilde{k}\sqrt{\frac{\eta}{\eta-1}}$, $\lambda_2=-1+\tfrac{\sigma-\gamma^2(\eta+2)}{2\eta\gamma}$. Plugging this ansatz into the radial equation yields a 3-term recurrence relation for the coefficients $a_n$:
	\begin{equation}
		\label{eq:Leaver_recurrence}
		\alpha_na_{n+1}+\beta_na_n+\delta_na_{n-1}=0,
	\end{equation}
	with $a_{-1}=0$, and normalization chosen such that $a_0=1$. The terms $\alpha_n,\beta_n$, and $\gamma_n$ are functions of $\omega$ and the other parameters, and explicitly read
	\begin{widetext}
	\begin{equation}
		\nonumber
		\begin{aligned}
			\label{eq:Leaver_coefficients}
			\alpha_n&=4\left(n+1\right)\eta^2(\eta-1)\left((n+1)\sqrt{\eta-1}+2\tilde{k}\sqrt{\eta}\right),\\
			\beta_n&=2\eta\bigg[2\gamma^4\eta^2\sqrt{\eta-1}-\sigma(\eta-1)\gamma\left((2n+1)\sqrt{\eta-1}+2\tilde{k}\sqrt{\eta}\right)-\gamma^2\eta\big(2\tilde{k}(1-\eta)\sqrt{\eta}(2+4n+3\gamma)\\
			&~~~~~~~~~~~-(\eta-1)^{3/2}\left(2+4\tilde{k}^2+2n(2+2n+3\gamma)+3\gamma+2\Lambda\right)+2\sigma\sqrt{\eta-1} \big)\bigg],\\
			\delta_n&=4\eta(\eta-1)^{3/2}\bigg[n\left(\gamma(\sigma-\gamma^2(\eta+2))-2\tilde{k}\eta\gamma^2\sqrt{\frac{\eta}{\eta-1}}\right)
            -\gamma^2\left(n^2\eta+\tilde{k^2}(\eta+1)\right)+\tilde{k}\gamma\sqrt{\frac{\eta}{\eta-1}}(\sigma-\gamma^2(\eta+2))\\
			&~~~~~~~~~~~~~~~~~~~~~~~~~~~~-\frac{(\sigma-\gamma^2\eta)((\eta-1)\sigma-\gamma^2\eta(\eta+3))}{4\eta(\eta-1)}\bigg].
		\end{aligned}
	\end{equation}
    \end{widetext}
	In the large-$n$ limit, the ratio of series coefficients in Eq.~\eqref{eq:Leaver_ansatz} behaves as $a_{n+1}/a_n\sim 1\pm\tfrac{\sqrt{2(\eta-1)\gamma/\eta}}{\sqrt{n}}+\mathcal{O}(n^{-1})$. Recalling that $\gamma>0$ for bound modes, the series converges uniformly if we pick the minus branch of the solution. This branch corresponds to minimal solutions of the radial equation~\cite{Leaver:1985ax}, with the ratio of successive $a_n$ coefficients given by the infinite continued fraction
	\begin{equation}
		\label{eq:continued_fraction}
		\frac{a_{n+1}}{a_n}=\frac{-\delta_{n+1}}{\beta_{n+1}-}\frac{\alpha_{n+1}\delta_{n+2}}{\beta_{n+2}-}\frac{\alpha_{n+2}\delta_{n+3}}{\beta_{n+3}-}...\,,
	\end{equation}
	where we have used standard notation for a continued fraction. From Eq.~\eqref{eq:Leaver_recurrence}, we can easily see that $a_1/a_0=-\beta_0/\alpha_0$. Taking this in tandem with Eq.~\eqref{eq:continued_fraction} yields the characteristic equation for the eigenfrequencies of the system:
	\begin{equation}
		\label{eq:characteristic_eqn}
		0=\beta_0-\frac{\alpha_0\delta_1}{\beta_1-}\frac{\alpha_1\delta_2}{\beta_2-}\frac{\alpha_2\delta_3}{\beta_3-}....
	\end{equation}
	We may also invert Eq.~\eqref{eq:characteristic_eqn} an arbitrary number of times to obtain the equivalent relation
	\begin{equation}
    \begin{aligned}
		\label{eq:characteristic_eqn1}
		\left(\beta_n-\frac{\alpha_{n-1}\delta_{n}}{\beta_{n-1}-}...\frac{\alpha_{0}\delta_{1}}{\beta_{1}}\right)=\frac{\alpha_{n}\delta_{n+1}}{\beta_{n+1}-}\frac{\alpha_{n+1}\delta_{n+2}}{\beta_{n+2}-}...\,,
        \end{aligned}
	\end{equation}
	relating a finite continued fraction on the left hand side to an infinite continued fraction on the right hand side. Normal mode solutions are found by numerically finding the $\tilde{\omega}$ values which satisfy Eq.~\eqref{eq:characteristic_eqn1}, and their corresponding wavefunctions are found by generating series coefficients via Eq.~\eqref{eq:continued_fraction}.
	
\bibliography{Refs}

\end{document}